\documentclass{twocol}

\usepackage{graphicx}
\usepackage{float}
\usepackage{amssymb,amsfonts,amsmath}
\usepackage{color}
\usepackage{subfigure}
\usepackage[bbgreekl]{mathbbol}
\usepackage{bm}
\usepackage{soul}

\def \Ca{C_\alpha}

\url{www.pnas.org/cgi/doi/10.1073/pnas.0709640104}
\copyrightyear{2008}
\issuedate{Issue Date}
\volume{Volume}
\issuenumber{Issue Number}

\begin{document}


\title{SOD1 Exhibits Allosteric Frustration to Facilitate Metal Binding Affinity}

\author{Atanu Das\affil{1}{Department of Physics and Astronomy, University of British Columbia, Vancouver,
Canada}
\and
Steven S. Plotkin\affil{1}{}}

\contributor{Submitted to Proceedings of the National Academy of Sciences
of the United States of America, Sept 2012}

\maketitle

\begin{article}
\begin{abstract} 
Superoxide dismutase-1 (SOD1) is a ubiquitous, Cu and Zn binding, free
radical defense enzyme whose misfolding and aggregation play a
potential key role in amyotrophic lateral sclerosis, an invariably
fatal neurodegenerative disease. 
Over 150 mutations in SOD1 have been identified with a familial form
of the disease, but it is presently not clear what unifying features,
if any, these mutants share to make them pathogenic.  Here, we develop
several new computational assays for probing the thermo-mechanical
properties of both ALS-associated and rationally-designed SOD1variants.
Allosteric interaction free energies between residues and metals are
calculated, and a series of atomic force microscopy experiments are
simulated with variable tether positions, to quantify mechanical
rigidity ``fingerprints'' for SOD1 variants. 
Mechanical fingerprinting studies of a series of C-terminally
truncated mutants, along with an analysis of equilibrium dynamic
fluctuations while varying native constraints, potential energy change
upon mutation, frustratometer analysis, and analysis of the coupling
between local frustration and metal binding interactions for a glycine
scan of 90 residues together reveal that the apo protein is internally
frustrated, that these internal stresses are partially relieved by
mutation but at the expense of metal-binding affinity, and that the
frustration of a residue is directly related to its role in binding
metals. 
This evidence points to apo SOD1 as a strained intermediate with
``self-allostery'' for high metal-binding affinity. 
Thus, the prerequisites for the function of SOD1 as an antioxidant
compete with apo state thermo-mechanical stability, increasing the
susceptibility of the protein to misfold in the apo state. 
\end{abstract}


\abbreviations{(F,S)ALS, (familial, sporadic) amyotrophic lateral sclerosis; WT,
  wild-type; ZBL, Zn-binding loop;  
ESL, electrostatic loop; RMSF, root mean squared fluctuations; SASA,
solvent-accessible surface area; PTM, post-translational modification;
Cu,Zn(SS), holo, disulfide present; E,E(SH), apo, disulfide-reduced;
WHAM, weighted histogram analysis method; SI, Supporting Information Appendix}

\dropcap{A}llosteric regulation canonically involves the modulation of a
protein's affinity for a given ligand A through the binding of a
separate ligand B to a distinct spatial location on the protein. The
modulatory binding site at the distinct location is referred to as the
allosteric site, and the interaction between the allosteric site and the
putative agonist binding site is referred to as an allosteric
interaction.  Early models of allostery were used to explain ligand
saturation curves for hemoglobin in terms of subunit interactions that
would induce binding
cooperativity~\cite{MonodJ65,KoshlandDE1966comparison}.
Cooperativity may be quantified through the non-additivity in the
binding energies of ligand and allosteric effector~\cite{WeberG75}. More
recently, allostery has been thought to be a more generic property
present even in single domain proteins~\cite{GunasekaranK04}. In this
context, positive cooperativity may be modulated 
through frustrated intermediates~\cite{FerreiroDU11}, 
which may enhance the conformational changes observed upon ligand
binding for allosteric proteins~\cite{Daily2007local}. 
In a single domain protein, the notion of
an allosteric effector can generalized to intrinsic protein side
chains that can enhance protein function or functionally important
motion at the expense of native stability- a kind of
``self-allostery''. Some allosteric activation mechanisms involve
mediation of conformational switches by non-native intra-protein
interactions~\cite{Gardino2009transient}, an effect predicted by energy
landscape approaches wherein barriers between conformational states
are “buffed” to lower energies~\cite{PlotkinSS03}. The potential
for novel allosteric regulators may vastly broaden candidate targets
for drug discovery~\cite{Christopoulos2002allosteric}. The internal
frustration required for cooperative allosteric 
function may have deleterious consequences however, if protein
stability is sufficiently penalized in intermediate states to enhance
the propensity for misfolding and subsequent aberrant oligomerization,
processes known to be involved in neurodegenerative
disease~\cite{ChitiF06}. Here we show that the ALS-associated protein  
Cu, Zn superoxide dismutase (SOD1)
is embroiled in such a conflict between stability and function. 

SOD1 is 
a homo-dimeric antioxidant enzyme of 32kDa, 
wherein each
monomer 
contains 153 amino acids, 
binds one Cu and one Zn ion,
and  
consists largely of an eight stranded greek key $\beta$
barrel with  
two large, functionally important
loops~\cite{TainerJA82,BertiniI98,ValentineJS05}. 
Loop VII or the electrostatic loop (ESL, 
residues 121-142) 
enhances the enzymatic activity of the protein by
inducing 
an electrostatic funnel towards
a redox active site centered on the Cu ion~\cite{GetzoffED92}. 
Loop IV or the Zn-binding loop (ZBL 
residues 49-83), contains
histidines H63, H71, and H80 
as well as D83, which 
coordinate the Zn ion and, along with a disulfide bond between C57 and
C146, enforce concomitant tertiary structure in 
the protein. 
The Cu on the other hand is 
coordinated by 
H46, H48, and H120, and the 
bridging histidine H63 between the Cu and Zn, which  
are located primarily in 
the Ig-like  
core of the protein. 

SALS inclusions are immunoreactive to misfolding-specific SOD1
antibodies~\cite{BoscoDA10,ForsbergK10}. 
Such misfolded aggregates
may be initiated from locally (rather than globally) unfolded states
that become accessible, for example, via thermal fluctuations or rare
events~\cite{ChitiF09}, e.g.
near-native aggregates or 
aggregation precursors were 
found for an obligate
monomeric SOD1 variant~\cite{NordlundA06}, 
and for FALS mutants S134N~\cite{ElamJS03nsb,BanciL2005} and E,E(SS)
H46R~\cite{ElamJS03nsb}. 
These above studies have motivated the present computational study,
which focuses on the native and near-native thermo-mechanical properties of
mutant, WT and post-translationally modified SOD1.

\section{Results}
\subsection{Cumulative distributions of work values can discriminate SOD1
  mutants and premature variants.}
\label{apoG127X}

Mechanical probes were employed by simulating tethers on the residue
closest to the center of mass of the SOD1 variant, and on various
residues on the protein surface.
The experimental analogue

\begin{figure*}[h]
\centerline{\includegraphics[width=.75\textwidth]{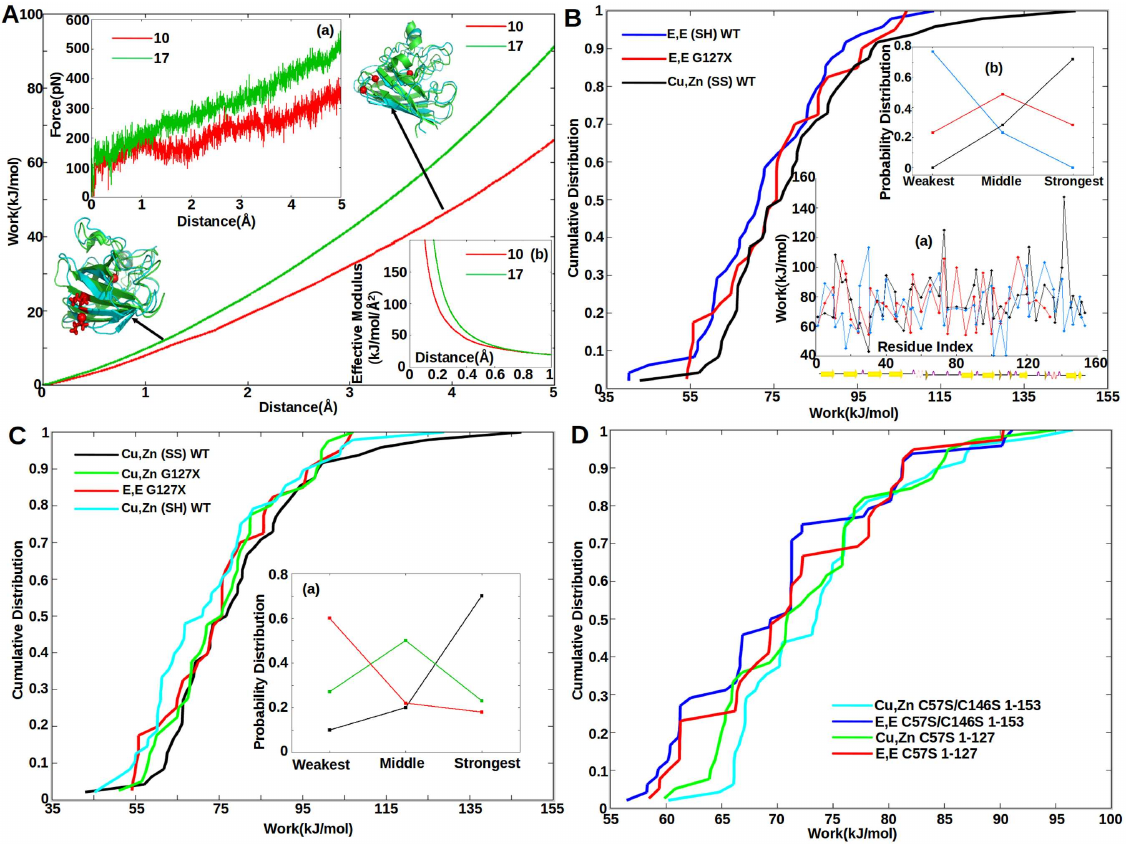}}
\caption{
(Panel A)
Force (inset a), Work (main panel), and effective modulus (inset
b) as a function of extension.
Tethers are placed at
the C$_\alpha$ atom closest to the center of mass of the SOD1 monomer
(H46), and the C$_\alpha$ atom of either residues G10 (red) or I17
(green), in separate pulling assays.
Ribbon representations of the protein are also shown; the
tethering residue is shown in licorice rendering (in red) and the
center C$_\alpha$ as a red sphere. The initial equilibrated (at
$0$~\AA~, green ribbon) and final (at $5$~\AA~, blue ribbon) structures are aligned to each other by
minimizing RMSD.
(Panel B)
(Inset a) Work profiles of Cu,Zn(SS) WT (black), E,E(SH) WT
(blue), and E,E G127X SOD1 (red) vs. sequence index. Secondary
structure schematic is shown underneath.
(Main panel) Cumulative distributions of the work values in inset (a).
E,E G127X is more stable than full-length E,E (SH) SOD1 (p=9e-7).
(Inset b) Fraction of the 48 incidences that each variant had either
the weakest, strongest, or middle work value, e.g. E,E(SH) SOD1 is
weakest 80\% of the time and is never the strongest variant.
(Panel C)
Cumulative work distributions for Cu,Zn (SS) WT
(black), Cu,Zn  G127X (green), E,E G127X (red), and Cu,Zn (SH) WT (cyan).
Cu,Zn G127X  is destabilized with respect to full-length Cu,Zn (SS) WT (p=6.2e-8).
(Inset a) Same analysis as panel B, inset b, for the variants
Cu,Zn(SS) WT, Cu,Zn G127X, and E,E G127X.
(Panel D)
Cumulative distributions for serine mutant SOD1 variants
demonstrate that C-terminal truncation
stabilizes the apo form but destabilizes the holo form (see text).
}
\label{figFvsx}
\end{figure*}

to such an {\it in silico} approach would
require multiple AFM or optical trap assays involving numerous residue pairs
about the 
protein surface as tethering points. This is difficult and time-consuming
to achieve in practice, which thus provides an opportunity for the present
simulation approaches.
All proteins in this study were taken in the monomeric form- many of
them such as E,E(SH) SOD1, G127X, and G85R  
are either naturally 
monomeric or have significantly reduced dimer stability. 
The ALS-associated truncation mutant G127X contains a frameshift
insertion which results in 6 non-native amino
acids following Gly127, after which a truncation sequence terminates
the protein 20 residues short of the putative
C-terminus~\cite{JonssonPA04,GradLI11}. 
Here 
PTMs refer to 
processes involved in the {\it in vivo}
maturation of SOD1, 
including disulfide bond formation, and 
metallation by Cu and Zn. 

The simulated force-extension profile may be used to obtain the work
required to pull a given residue to a given distance
(Fig. \ref{figFvsx}A). In this assay, we found that large
effective stiffness moduli were observed for small perturbing
distances (Fig. \ref{figFvsx}A inset b). These were attributable primarily to side chain-side chain
``docking'' interactions in the native structure
(SI Appendix, Fig. S2).
The work to pull a given residue to a distance sufficient to
constitute an anomalously large fluctuation, e.g. 5\AA, can be
calculated as a function of sequence index for a given SOD1
variant, resulting in a characteristic mechanical profile or
``mechanical fingerprint'' for that protein (Fig.~\ref{figFvsx}B,
inset a). 
The work values at 5\AA~do not correlate with RMSF
values of the corresponding residues~\cite{DasA12jmb}. 
A representative subset of 48 residues was obtained as
described in the SI. 
We found that such a profile was independent of how the
protein was initially constructed before equilibration, i.e. from what
PDB, but was clearly different between WT SOD1 and mutants such as
G127X, as well as other PTM
variants such as E,E(SH) 
SOD1 (SI Fig.~S4).  Table S1 in SI gives the
correlation coefficients between variants.
A given PTM globally modulates the mechanical profile, inducing both
local and non-local changes in stability that can be both
destabilizing

\begin{figure*}[h]
\centerline{\includegraphics[width=.75\textwidth]{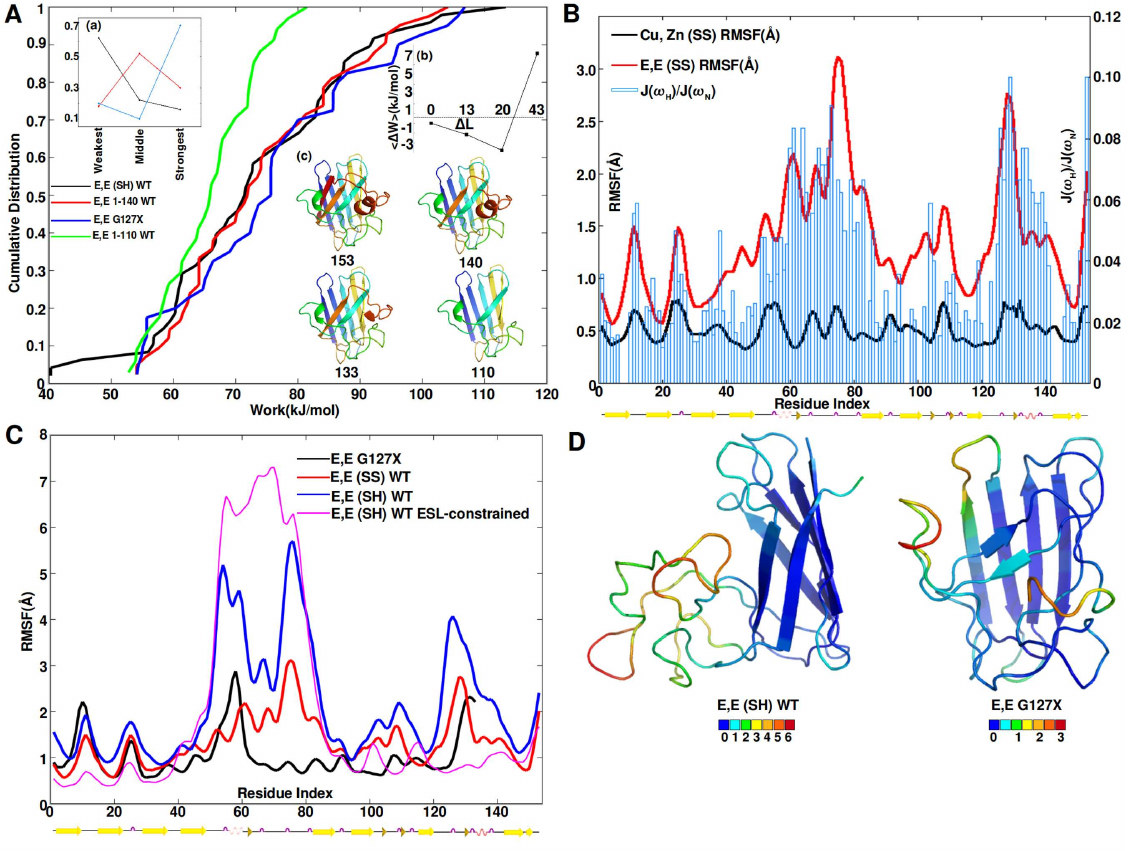}}
\caption{
(Panel A) Cumulative distributions of work values for C-terminal-truncated SOD1
variants of variable sequence length
show a non-monotonic trend in mechanical stability.
All variants are metal-depleted and have no disulfide bond. Sequences
are given in the legend (see text).
(Inset a) Comparing the cumulative distributions in the main panel, that of the mutant
G127X is most commonly the strongest, full-length SOD1 is most
commonly the weakest, and 1-140 is most often in
the middle. (Inset b) Change in work value
  $\overline{W}_{\tiny{MUT}}-\overline{W}_{\tiny{WT}}$ averaged over
  residues, as a function
truncation length, for the SOD1 variants in the main panel.
(Inset c) Ribbon schematics of the various truncation mutants, colored
blue to red from N- to C-terminus, labeled by C-terminal residue.
(Panel B) Simulated native-basin dynamical fluctuations (RMSF) in explicit SPC solvent, for
Cu,Zn(SS) (black) and E,E(SS) SOD1 monomer (red), along
with the experimentally measured ratio of spectral density functions
$J(\omega_H)/J(\omega_N)$ of obligate monomeric E,E(SS) F50E/G51E/E133Q SOD1 (blue
bars)~\cite{BanciL03}. Correlation coefficient is $r=0.78$.
(Panel C)
Simulated RMSF for SOD1 variants E,E G127X (black), E,E(SS) (red),
E,E(SH) (blue), and E,E(SH) with the ESL
constrained to be natively structured (magenta). The
presence of native stress is indicated by the increased disorder of
the ZBL upon structuring the ESL (see text).
(Panel D) Snapshots of typical structures of E,E(SH) and G127X
SOD1 from equilibrium simulations, color coded by the mean RMSF for
each residue; RMSF increases from blue to red according to the scale bars shown.
}
\label{figrmsfsasa}
\end{figure*}

in some regions and stabilizing in others. 
A detailed study
of the mechanical consequences of mutations or PTMs, 
e.g. by studying the
long-range communication through interaction networks in the
mutant vs WT protein (see e.g.~\cite{KhareSD06pnas,PotterSZ07,SchuylerAD11,EdwardsSA12})
is an interesting topic of future research.

We also found that the mechanical work profile was nearly independent
of whether an explicit or implicit solvent model was used in the
simulations, even though native basin fluctuations
(RMSF) showed significant scatter in comparing implicit and explicit solvent
models (SI Fig.~S5). Equilibrium fluctuations are
much more sensitive to solvent
models than the large scale perturbations we consider here.
On the other hand, a structure-based G\={o} model
does only a modest job of capturing the mechanical profile, and is
poor in particular for
residues where electrostatic interactions play a significant role in
stability (SI Fig.~S6). 
Interestingly, the default energy scale of 1 kJ/mol
for each contact or dihedral interaction
used in current G\={o} models~\cite{WhitfordP09}
captures the overall energy scale of the work profile quite well for
SOD1, though it must likely be modified for other proteins such as
thermophilic proteins. 

In support of the thermodynamic relevance of a SOD1 variant's
mechanical fingerprint, we found that
the non-equilibrium work values obtained by pulling a residue to 5\AA~
strongly correlate with the equilibrium free energy change for the
same process, as calculated using the 
WHAM method 
(r=0.96, SI Fig.~S7). Thus, the work profiles
obtained are an accurate measure of the corresponding 
local thermodynamic stability profile, up to a scaling factor.


Comparing the mechanical fingerprints of E,E(SH) SOD1 and that of
G127X (Fig.~\ref{figFvsx}B, inset a), we see that many discrepancies between SOD1 variants
and WT are difficult to disentangle solely from the work profile, or even
from histograms of the work values (SI Fig.~S3). 
However, the mechanical discrepancies
emerged quite naturally from the cumulative distribution of work values.
Mechanical scans were thus used to construct the cumulative
distributions, which then allowed us to distinguish
stabilizing energetics in various forms of PTM SOD1, e.g. between
E,E(SH), G127X, and Cu,Zn(SS) SOD1 in Figure~\ref{figFvsx}B.
Not all residues needed to be sampled to obtain the
cumulative distribution- we found convergence to within $\approx 1$ kJ/mol
after a sample size of about 40 residues
(SI Fig.~S8). 
This does not mean the
mechanical profile obtained from $48$ residues is representative of
the stability in the corresponding regions of the protein: the
mechanical work values were essentially uncorrelated for amino acids about 3
residues apart.

Comparison of the cumulative work distributions shows that the
ALS-associated mutant Cu,Zn(SS) A4V is slightly more mechanically malleable
than Cu,Zn(SS) WT, particularly in the weaker regions of the protein 
(SI Fig.~S13). 
However the mechanical weakening due to mutation is substantially larger in the E,E(SS) state,
and largest in the E,E(SH) state (green curves in SI Fig.~S13). 
Experimental measurements of melting temperature also have shown
increased susceptibility to mutation in the E,E(SH) state for
A4V~\cite{FurukawaY05}. 
We also note from SI Figure~S13 that disulfide reduction in the WT apo state, E,E(SS)
$\!\!\rightarrow\!\!$ E,E(SH), actually increases the rigidity of the
native basin~\cite{DasA12jmb}, even though the net thermodynamic stability is decreased,
primarily due to increased unfolded entropy~\cite{HornbergA07}.

\subsection{Lack of post-translational modifications mechanically
  destabilizes SOD1, however the truncation mutant G127X stabilizes the
  apo, disulfide-reduced protein.}
\label{apoG127X}

The cumulative distribution of E,E(SH) lies to the left of that for Cu,Zn(SS) SOD1
in Figure~\ref{figFvsx}B for all rank-ordered work values,
illustrating that that variant is more malleable and thus more susceptible to
perturbing forces that might induce the conformational changes
accompanying misfolding. Likewise, Cu,Zn(SH) is destabilized with
respect to Cu,Zn(SS) in
Figure~\ref{figFvsx}C, and the other PTM variants in SI Figure~S14
show that 
lack of PTMs reduces the mechanical rigidity.

On the other hand, the E,E G127X truncation mutant, which lacks both
metals, a disulfide bond, and part of the sequence, shows 
{\it increased} mechanical stability over E,E(SH) WT ($p = 9$e-7, Fig.~\ref{figFvsx}B).
Comparing the profiles, E,E(SH) is most commonly the weakest, Cu,Zn(SS) is most
commonly the strongest, and G127X is most commonly in the middle
(Fig.~\ref{figFvsx}B Inset b). 
Apparently the C-terminal region of the protein mechanically stresses the remainder
of the protein {\it when PTMs are absent}, reducing its mechanical
stability.  However, a metastable holo variant of G127X containing
metals in their putative positions
is {\it less} mechanically stable than holo WT ($p=6.2$e-8, 
Fig.~\ref{figFvsx}C). 
In the full-length holo protein,
native stabilizing interactions are more apt to be minimally
frustrated, 
and thus C-terminal truncation induces
softening of the native structure.
The metallation of G127X only marginally stabilizes it
($p=6.7$e-3, Fig.~\ref{figFvsx}C inset a), while metallation of WT
protein results in substantial mechanical stabilization
(SI Fig.~S14). 
G127X lacks C146, so Cu,Zn(SH) WT may be a more appropriate
comparison than Cu,Zn(SS) WT. 
Cu,Zn(SH) SOD1 is less stable than either Cu,Zn G127X  
or E,E G127X (Fig.~\ref{figFvsx}C),
however Cu,Zn(SH) contains a protononated Cysteine C146, 
which is destabilizing, and not present in G127X.

To deconvolute the effects of protonated Cysteines, we examined a set of 4 serine mutant proteins:
Cu,Zn C57S/C146S 1-153 (full length), E,E C57S/C146S 1-153, Cu,Zn C57S
1-127 (truncated at residue 127), and E,E C57S 1-127. For this system
of proteins, Cu,Zn C57S/C146S 1-153 is clearly the most stable over
most of the range of work values
(Fig.~\ref{figFvsx}D), and Cu,Zn C57S 1-127 is destabilized with
respect to it. E,E C57S 1-127 is further destabilized with respect to
its holo form, but is {\it more} stable than the full length apo form E,E
C57S/C146S 1-153. This demonstrates that C-terminal truncation
stabilizes the apo form but destabilizes the holo form. 
This conclusion is robust to changes in pulling distance (SI Fig.~S15).

\subsection{Short length C-terminal truncation mechanically stabilizes the apo
  protein, while sufficiently long C-terminal truncation destabilizes
  it.}
\label{ctermtrunc}

The mechanical properties of two additional truncated constructs, 
E,E 1-140 (WT sequence), and E,E 1-110, 
were assayed to investigate the crossover from increased to eventually
reduced mechanical stability as the length of truncation is increased
(Fig.~\ref{figrmsfsasa}A inset b).
All truncation variants are missing the putative disulfide bond.
Figure~\ref{figrmsfsasa}A shows cumulative distributions for the above
variants along with E,E(SH) SOD and E,E G127X. The truncation E,E 1-140
mechanically stabilizes E,E(SH) SOD1 ($p=
1$e-3), and G127X is further stabilized with respect
to 1-140 ($p= 9$e-4).
Comparison of the triplets of rank-ordered work values corresponding
to E,E(SH), E,E 1-140, and E,E G127X supports this conclusion
(Fig.~\ref{figrmsfsasa}A  inset a).
We have omitted the 3
least stable work values for all variants in calculating statistical
significance- these are outliers for E,E(SH) that would dominate the
result towards the 
conclusion that we have arrived at without their inclusion.
Variant E,E 1-110 has significantly compromised mechanical rigidity
and may not be thermodynamically stable- the mechanical assay only
probes malleability of the native basin.



\subsection{Simulated fluctuations correlate with experimental
  spectral density functions for apo SOD1.}  

We performed 20ns equilibrium simulations in explicit solvent (SPC
water model) for Cu,Zn(SS) and E,E(SS) SOD1.
Native-basin RMSF values show significant scatter between explicit and
implicit solvent models compared to mechanical work values (SI Fig.~S5),
so for analysis of equilibrium fluctuations, explicit solvent is used. 
Figure~\ref{figrmsfsasa}B plots the resulting equilibrium RMSF
for Cu,Zn(SS) and E,E(SS) SOD1; SI Figure~S9 also gives the
solvent-accessible surface area of backbone amide Nitrogens (SASA${}_N$) for
the same equilibrium trajectory. 
The main effect of metal loss is to
induce solvent exposed, disordered, and dynamic Zn-binding and
electrostatic loops (loops IV and VII). A
moderate increase in dynamics is observed for loop VI as well.
The preferential increase in dynamics of the ZBL and ESL upon
  metal loss is
consistent with experimental measurements of
the ratio of spectral density functions $J(\omega_H)/J(\omega_N)$- 
a measure of dynamics fast compared to the tumbling rate
(correlation coefficient $r=0.78$). 
The experimental measurements of spectral density~\cite{BanciL03} 
are obtained for a monomeric E,E(SS) SOD1
mutant F50E/G51E/E133Q~\cite{BertiniI94}. 

\subsection{Dynamic fluctuations in C-terminal truncation mutants reveal native frustration in
  apo SOD.}

Because G127X is more mechanically stable than E,E(SH) SOD1, native
basin fluctuations were investigated to see if the extra stability
of G127X was recapitulated in equilibrium dynamics. 
Figure~\ref{figrmsfsasa}C shows 
that indeed the RMSF are substantially enhanced
in E,E(SH) relative to G127X, in particular in the ZBL and ESL (loops IV
and VII). This is true even though $\beta$-strand 8, N-terminal to the
ESL, constrains E,E(SH) and is absent in G127X. Though more dynamic and
mechanically malleable,
E,E(SH) is not more solvent-exposed (by SASA${}_N$) than G127X, indicating a collapsed,
dynamic globule with non-native interactions (SI Fig.~S10). 
As well, E,E(SH) is more dynamic than E,E(SS) SOD1
(Fig.~\ref{figrmsfsasa}C), but less solvent exposed than E,E(SS),
particularly in the ZBL and ESL (SI Fig.~S10). Collapse and non-native
interactions are not hindering the dynamics of the ZBL and ESL;
snapshots from simulations for E,E(SH) WT and G127X are shown in Figure~\ref{figrmsfsasa}D. 
Similar condensation phenomena are
also observed in prion protein~\cite{LiL09} and
NHERF1~\cite{ChengH2009autoinhibitory}.
In a large scale study of 253 proteins across several fold
families~\cite{BensonNC08}, RMSF and SASA showed poor correlation
($r\approx 0.35$ for backbone carbons).


It is intriguing that the additional constraint of
structuring $\beta$8 in E,E(SH) SOD1 results in enhanced rather
than suppressed disorder in the ZBL.
One potential explanation is that because
there is more residues in the ESL present in E,E(SH) than in G127X,
the ESL forces the ZBL  to be more expanded by steric repulsion, and
thus more dynamic, i.e. a ``polymer brush''
effect. 
Another possible explanation is that the order induced by structuring
$\beta$8 in E,E(SH) SOD1 induces frustration and consequent
strain elsewhere in the protein, resulting in induced disorder in the
ZBL.
We differentiated these two scenarios by
applying native constraints between the $\Ca$ atoms in the ESL and
the rest of protein, but excluding contacts between the ESL and the
ZBL: harmonic springs were applied to all pairs of $\Ca$
atoms within 4\AA~in the native holo
structure that involved contacts either within residues 133-153, or
between residues 133-153 and either residues 1-40 or 90-153.
This procedure further constrains the ESL and $\beta$8 to be natively
structured, removing any polymer brush effect, but enhancing any
native strain. Consistent with a model involving native frustration,
the ZBL loop IV becomes {\it more} disordered in E,E(SH) SOD1 upon implementing ESL/$\beta$8 native
constraints (Fig.~\ref{figrmsfsasa}C).

\subsection{``Frustratometer'' results and potential energy changes support increased frustration
  in apo WT SOD1 with respect to apo ALS-associated mutants.}
\label{frustratometer}

\begin{figure}[h]
\centerline{\includegraphics[width=.5\textwidth]{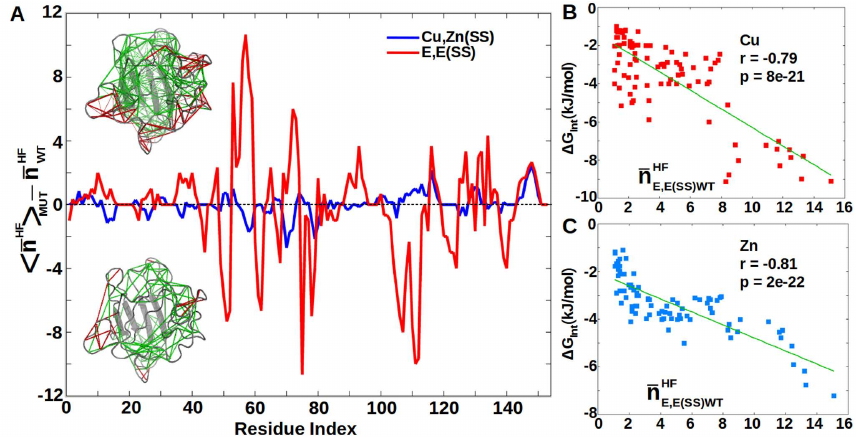}}
\caption{
(A) Frustrated contacts (in red) and unfrustrated contacts (in green)
for E,E(SS) WT SOD1, (B) Same contacts as (A) for the average over 22
ALS E,E(SS) mutants (see text).
(C) The mean number of frustrated contacts within a sphere of radius
5\AA~centered on each C${}_\alpha$ atom, is found as a function of residue
index. Ensemble averages are taken from 50 snapshots in an equilibrium
simulation. This is done for both the Cu,Zn(SS) state and the E,E(SS)
state, for both the  WT sequence, and for 22
mutant sequences. The 22 mutant sequences are averaged to
obtain the ensemble and mutant averaged number of contacts as a
function of residue index $i$, $\left< n^{\mbox{{\tiny HF}}}(i)\right>_{\mbox{{\tiny
        MUT}}}$. Plotted is the
difference between $\left< n^{\mbox{{\tiny HF}}}(i)\right>_{\mbox{{\tiny MUT}}}$ and the corresponding
numbers for the WT sequence $n_{\mbox{{\tiny WT}}}^{\mbox{{\tiny HF}}}(i)$ . A positive
number would indicate an increase in frustration upon mutation. Holo
state is shown in blue and has an average of +5 contacts; apo state is
shown in red and has an average of -22 contacts.
(B) Interaction free energy between a residue's side chain and the Cu
ion, plotted as a function of the E,E(SS) ensemble-averaged number of
highly-frustrated contacts that residue has (r=-0.79, p=8e-21).
(C) Same as in Panel B but for the Zn ion (r=-0.81, p=2e-22).
}
\label{figfrus}
\end{figure}

Typically frustrated contacts, according to the ``frustratometer'' method
developed by Wolynes and colleagues~\cite{FerreiroDU11}, were found by
averaging 50 snapshots from an equilibrium ensemble for WT SOD1, and
for each of 22 ALS-associated SOD1 mutants (SI Table~S3). 
Both E,E(SS), and Cu,Zn(SS) states were
analyzed. Results were averaged over the 22 mutants to yield the mean
number of frustrated contacts at a given residue position, for the
``average'' mutant. Taking the difference of this quantity with that
for WT SOD1 gives the mean change in frustration upon mutation, which 
increased for holo SOD1 by about $5$ total contacts, but {\it decreased}
for apo SOD1 by about $22$ total contacts (Fig.~\ref{figfrus}A). This result again supports
a frustrated apo state in SOD1. Direct computation shows on
  average about 38 more highly-frustrated contacts in E,E(SS) SOD1
  than  in Cu,Zn(SS) SOD1 (SI Fig.~S17).

\begin{figure}[h]
\centerline{\includegraphics[width=.5\textwidth]{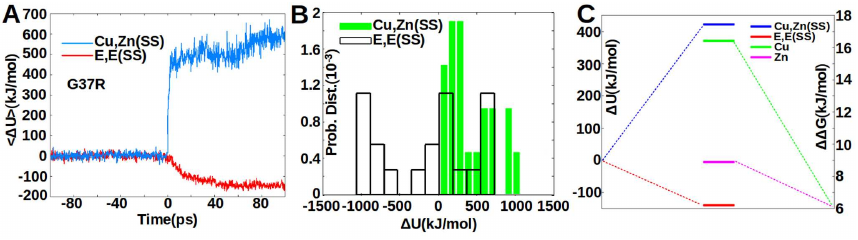}}
\caption{
(A) Change in potential energy $\Delta U(t)$ as a function of {\it in
  silico} time, before and after implementing
the mutation G37R. (B) Distribution of the asymptotic potential energy
change $\Delta U(\infty)$ for 22 ALS mutants (see text).
(C) Mean potential energy change averaged over mutants, for both the
holo state and the apo state, along with the mean difference,  WT
minus mutants, in both Cu
and Zn binding free energy.
}
\label{figDU}
\end{figure}

For the same set of mutants, we found that the 
ensemble-averaged total potential energy 
in the native state increased upon mutation for the Cu,Zn(SS)
protein, but decreased upon mutation in the E,E(SS) protein. This is
consistent with the above frustratometer results. 
The ``time-resolved''
change in potential energy $\Delta U$ upon {\it in silico} mutation for a representative
mutant (G37R) is shown in Figure~\ref{figDU}A. The distribution of
potential energy changes for the 22 ALS-associated mutants listed in
SI Table~S3 is shown in Figure~\ref{figDU}B, which shows potential energy
``cost'' upon mutation for Cu,Zn(SS) mutants, but a significant
shift towards negative stabilizing values for E,E(SS) mutants.
The same
conclusion is obtained from 30 non-ALS alanine mutants (SI Fig.~S18).
Figure~\ref{figDU}C shows that the net effect of mutation on the potential
energy is to increase it on average in the holo state, but to {\it decrease}
it on average in the apo state, indicating stabilization. Moreover, all ALS mutants facilitate
metal release~\cite{DasA12jmb};
the net effect of mutation
on the Cu and Zn binding free energies is to decrease both of
them. Thus mutations relieve stress in the apo
state, while inducing loss of metal binding function. 
It thus appears that 
apo SOD1 has evolved to have
high affinity for metals at the expense of native stability and
increased frustration. 



\subsection{Frustrating residues in the apo state are allosteric
  effectors, and positively modulate metal affinity in proportion to
  their frustration} 
\label{allostery}

To test the extent to which the residues 
facilitating metal binding are frustrated,
we have calculated
the interaction free energy of each WT residue with both Cu and Zn, by considering thermodynamic cycles~\cite{WeberG75} involving metallation and residue
``insertion'' from a glycine at the corresponding position
(e.g. G4A). The interaction free energy $G_{int}$ between each residue
side-chain and either Zn or Cu is given (here specifically for residue
Ala 4 with Zn) by
\begin{eqnarray}
G_{int} \!\!\!\!\! &=& \!\!\!\!\! \Delta G \left( G\!\!\rightarrow\!\! A,
  \mbox{O}\!\!\rightarrow\!\! Zn\right) - \Delta G \left( G\!\!\rightarrow\!\!
  A,\mbox{O}\right)  -  \Delta G \left( G,\mbox{O}\!\!\rightarrow\!\!
  \mbox{Zn}\right)  \nonumber
\\
&=& \!\!\!\!\! \Delta G\left(A,\mbox{O}\!\!\rightarrow\!\!
  \mbox{Zn}\right) - \Delta G \left(G,\mbox{O}\!\!\rightarrow\!\!
  \mbox{Zn}\right)
\end{eqnarray}
where $\Delta G\left(A,\mbox{O}\!\!\rightarrow\!\!  \mbox{Zn}\right)$
is the free energy change of Zn insertion when alanine is present at
position 4, and $\Delta G \left(G,\mbox{O}\!\!\rightarrow\!\!
  \mbox{Zn}\right)$ is the free energy change of Zn insertion when
glycine (no side chain) is present at position 4. The other residues,
and Cu interactions, are handled analogously (see SI Methods). 

Figures~\ref{figfrus}B,C plot the above interaction free energy of a
residue with Cu or Zn, {\it vs.} the E,E(SS) ensemble-averaged number
of highly-frustrated contacts that residue has. Data are obtained for
90 glycine mutants listed in SI Table~S3, for residues that had at least
one highly-frustrated contact (see SI). 
The values of $G_{int}$ for all 90 mutants listed in SI Table~S3 are
negative, indicating cooperative interactions, 
wherein the WT residue facilitates binding of the metal. 
Each of these residues can be thought of as allosteric effectors,
positively modulating affinity for either metal. 
For both Cu and Zn, the larger the degree of frustration in the apo
state, the larger the role that residue has in facilitating metal
binding. No such trend is seen for Cu,Zn(SS) SOD1 (SI Fig.~S18).
This result strongly 
supports a model
of the apo state as an 
allosteric intermediate designed for high 
metal binding affinity at the expense of structural stability.

\begin{figure}[h]
\centerline{\includegraphics[width=.5\textwidth]{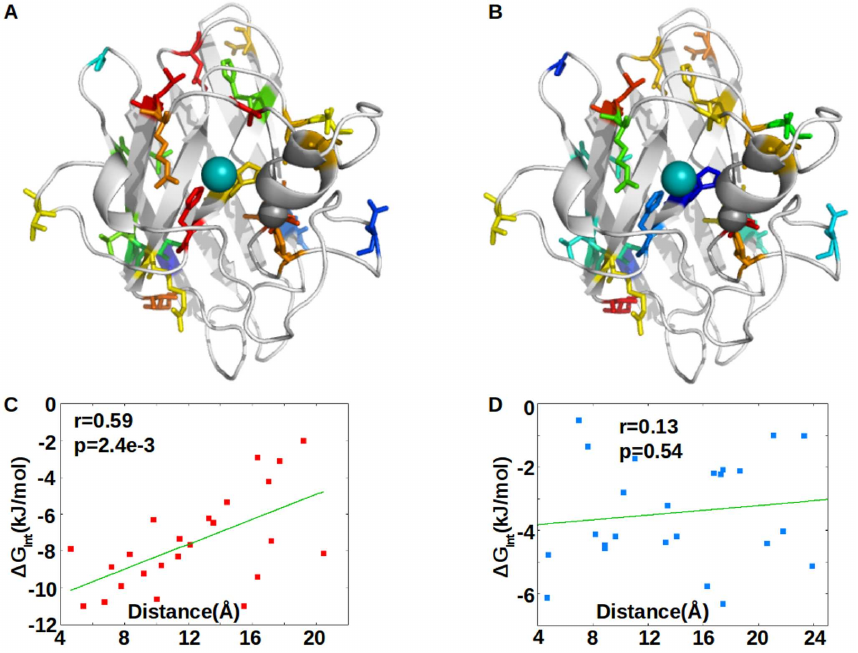}}
\caption{
Panel (A): Residues color-coded by interaction energy with the Cu ion
(depicted as a cyan sphere). The extent
of interaction is strongest in magnitude for red colored residues and
decreases to blue.
Panel (B): same as (A) for the Zn ion (depicted as a grey sphere).
Panel (C): Interaction energy with Cu correlates with the distance of the residue from the Cu
ion;  residues in close proximity more strongly interact.
Panel (D): Interaction energy with Zn does not correlate with distance
of the residue to the Zn ion, indicating non-local allosteric
effects.
}
\label{figinte}
\end{figure}

Finally, we test the distance-dependence of the interaction free
energy with the metals, for the set of 24 ALS-associated mutants in
SI Table~S3. 
In the case of Cu,
the allosteric regulation for metal affinity is significantly
correlated with the proximity of the residue to the Cu binding site
(Fig.~\ref{figinte}A,C). Interestingly, the allosteric regulation for
Zn binding is uncorrelated with distance to the Zn binding site and
thus nonlocal. This finding is consistent with experimental results
that Zn binding is concomitant with large structural change (partial folding)
of the protein~\cite{RobertsBR07}. Long-range coupling to the
Zn-binding region has also been observed in G93A
SOD1~\cite{Museth2009associated}.  
The same 
conclusion
is obtained for the 90 frustrated residues
given in SI Table~S3 (SI Fig.~S19).

\section{Discussion}
\label{discuss}

We have found here a connection between the allosteric design of
residues in premature superoxide dismutase towards high metal-binding
affinity, and the consequent frustration in the apo state of the
protein. 
A variety of results supported this conclusion. Mechanical profiles
were obtained from {\it in silico} AFM assays with variable tether
positions; in this context it was found that the C-terminal truncation
mutant G127X had higher mechanical rigidity than E,E(SH) SOD1,
implying the release of internal stresses upon removal of part of the
protein. Large truncation lengths eventually destabilized the protein.

The higher malleability of E,E(SH) SOD1 over G127X is recapitulated by
larger equilibrium dynamical fluctuations in the native
basin. Constraining loop VII (the ESL) to be natively structured only
increases fluctuations in loop IV (the ZBL), ruling out polymer brush
effects and supporting the native stress hypothesis. 

Implementing Wolynes's ``frustratometer'' method~\cite{FerreiroDU11}
shows that, perhaps surprisingly, more frustration is present on average in the
WT apo state than is present for apo mutants. For the holo state the
situation is reversed however, which is consistent with the notion that mutants
facilitate metal release. In fact every ALS-associated mutant we
have studied lowered the affinity for both Cu and
Zn~\cite{DasA12jmb}. We have found here that while these mutants
raise the potential energy of the holo state, they tend to lower the
potential energy of the apo state and thus stabilize it, consistent with frustratometer
results. 
A general comparative analysis of the decrease in potential energy and frustration
for apo ALS-associated mutants, along with the results from their individual
mechanical scans which generally show weakening with respect to local
perturbations, is an 
interesting topic for future work. SI Figure~S20 
compares the relevant quantities for the ALS mutants A4V and G127X.

Residues in apo SOD1 can be thought of as allosteric effectors
for metal binding. By considering the cooperativity in thermodynamic
cycles involving mutation to glycine and metal release, we quantified
the interaction free energies between residues in the protein 
and either Cu or Zn. All interaction
energies were negative, indicating positive modulation of metal
affinity, Moreover we found that function frustrates stability:
the stronger the interaction energy, the more frustrated the residue.
For Cu, the strength of the interaction significantly
correlates with proximity to the binding site. For Zn however, there is
no correlation with proximity, indicating a non-local allosteric
mechanism involving propagation of stress-release throughout the
protein, and consistent with the large structural changes accompanying
Zn binding. 

The above evidence 
points towards a paradigm wherein 
sequence evolution towards high metal affinity results in a trade-off
for significant native frustration in the apo state of SOD1.
A similar conclusion has been reached from studies of a SOD1
variant with Zn-coordinating ligands H63, H71, H80, and D83 mutated
to S, which in the apo form is stabilized with respect to E,E WT
SOD1~\cite{NordlundA2009}.
In this context, 
the C-terminal truncation in G127X can be
seen as an allosteric inhibition mechanism to Zn binding, in that
frustration is relieved in the apo native state, but Zn-binding
function is lost.
A similar scenario is observed for select mutants of subtilisin, a
serine protease whose function is regulated by Ca${}^{2+}$ binding. In
this protein, the mutation M50F preferentially stabilizes apo
subtilisin relative to the holo form, while weakening calcium binding
and promoting inactivation in the holo form~\cite{BryanPD00}.  

Native frustration in apo SOD1 as a result of allosteric cooperativity
in metal binding has potential consequences for the misfolding of
SOD1. The premature protein, or a protein that perhaps due to an
external agent has lost its metals, would show decreased thermal
stability relative to one that had not undergone sequence evolution
for high metal affinity.  In this sense, the tight binding of Zn and Cu essential for
enzymatic function
of the mature protein as an antioxidant puts the premature form in
additional peril for misfolding.


\begin{materials}
A full description of the methods is given in the
SI. 
Missense and truncation mutants of SOD1, both
ALS-associated and rationally designed, were equilibrated and used for
mechanical force, dynamic fluctuation, frustratometer, potential
energy, and WHAM metal affinity assays. 
Rationally-designed truncation and missense mutants studied here
include C57S/C146S, C57S 1-127, and WT
sequences 1-110 and 1-140. 
Frustration and metal-binding allostery assays used either
22 and 24 ALS-associated mutants respectively, or 90 glycine mutants
(SI Table~S3).
Mechanical profiles are obtained after 20ns pre-equilibration from steered MD
simulations (tether speed 2.5mm/s) in GBSA solvent with OPLS-AA/L
force field parameters. 
Robustness checks are shown in SI Figure~S15.
Monte Carlo methods yield the statistical signifance (error {\tiny 
$\approx$} 2.7kJ/mol, SI Fig.~S21). 
Fluctuation analysis used SPC explicit
solvent. Metal binding free energies are found from WHAM including
post-relaxation and validation by thermodynamic cycles. Frustration
calculations 
include protein conformations and protein-protein contacts only;
  i.e. metals are implicit in determining
  protein conformation but metal-protein interactions are not 
  explicitly included. Frustrated contacts were calculated using the frustratometer 
server http://frustratometer.tk. 


\end{materials}

\begin{acknowledgments}
We thank Neil Cashman, Will Guest, Ali Mohazab, Eric Mills, Paul Whitford,
and Stephen Toope for helpful and supportive
discussions. We acknowledge funding from PrioNet Canada, NSERC funding
to defray page charge costs, and we
acknowledge computational support from the WestGrid high-performance
computing consortium.
\end{acknowledgments}


\end{article}

\end{document}




\title{SOD1 Exhibits Allosteric Frustration to Facilitate Metal
  Binding Affinity  (Supporting Information)}

\author{Atanu Das\affil{1}{Department of Physics and Astronomy, University of British Columbia, Vancouver,
Canada}
\and
Steven S. Plotkin\affil{1}{}}

\contributor{Submitted to Proceedings of the National Academy of Sciences
of the United States of America, Sept 2012}

\maketitle

\begin{article}

\section{SI-text}

\subsection{Work-extension profiles provide a measure of local
  mechanical stability, and have distance-dependent stiffness moduli}
\label{workresults}

We performed pulling simulations on residues taken from the mid-points
of the protein sequences of superoxide dismutase predicted to be
either weak (referred to here as candidate epitopes) or strong (candidate anti-epitopes)
thermodynamically (see Fig.~\ref{figtabepitopes} 
).

\renewcommand{\thefigure}{S\arabic{figure}}
\renewcommand{\arraystretch}{1}
\begin{figure}[h]
\centerline{\includegraphics[width=0.4\textwidth]{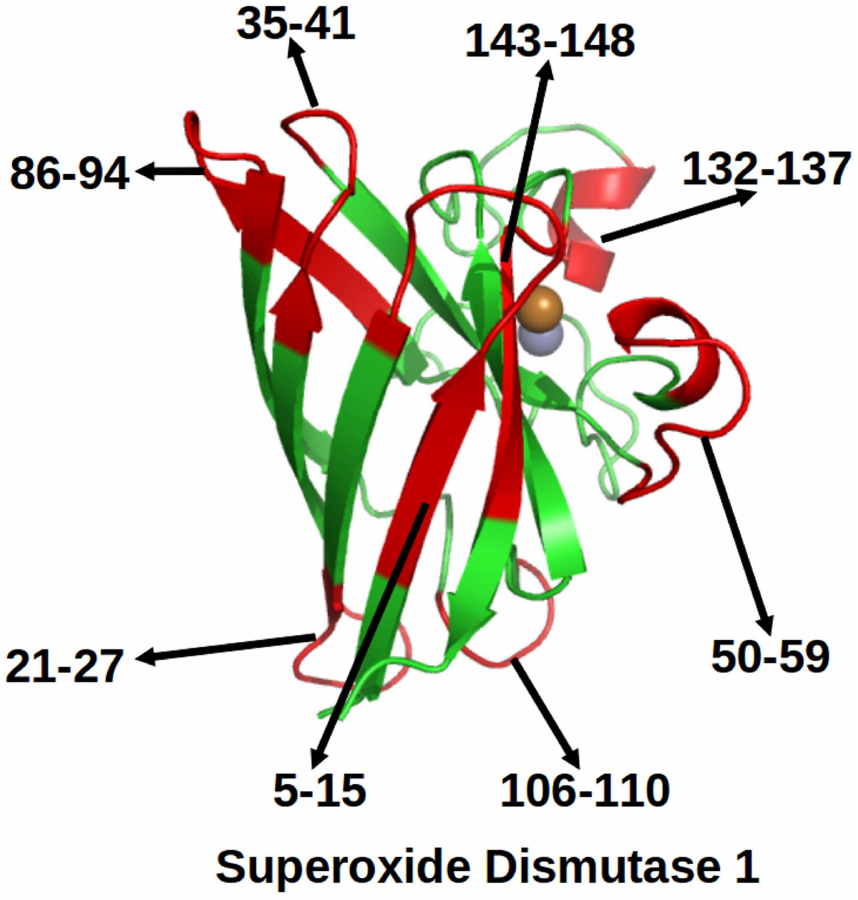}}
\bigskip
\begin{tabular}{ccc}
\hline
(Anti)-Epitope sequence & Center residue taken & Work(5 \AA) \\
\hline
$5-15$ & $10$ & $66.1$ \\
$(16-20)$ & $17$ & $91.5$ \\
$21-27$ & $24$ & $58.8$  \\
$(28-34)$ & $31$ & $66.3$  \\
$35-41$ & $38$ & $67.1$  \\
$(42-49)$ & $45$ & $83.4$ \\
$50-59$ & $54$ & $85.0$  \\
$(60-85)$ & $73$ & $125.0$ \\
$86-94$ & $90$ & $87.9$  \\
$(95-105)$ & $100$ & $97.8$  \\
$106-110$ & $108$ & $69.2$  \\
$(111-131)$ & $121$ & $113.6$  \\
$132-137$ & $135$ & $79.4$  \\
$(138-142)$ & $140$ & $99.7$  \\
$143-148$ & $145$ & $72.8$  \\
$(149-153)$ & $151$ & $76.6$  \\
\hline
\end{tabular}
\caption{Ribbon representation of monomeric SOD1 structure with
Cu and Zn metals shown as orange and gray spheres
respectively. Candidate misfolding-specific epitopes as predicted by the algorithm of Guest, Cashman
and Plotkin~\cite{GCPalgorithm09} are colored red, and their residue
numbers are indicated. In the Table - Epitopes, anti-epitopes, pulling residues, and resulting work
values for Cu,Zn (SS) WT SOD1.}
\label{figtabepitopes}
\end{figure}

A tethering point was
placed on the C$_\alpha$ atom at the center residue of a candidate epitope or
anti-epitope (see 
Methods below), and another tethering
point was placed on the  C$_\alpha$ atom closest to the center of mass
of the protein (histidine 46).
A plot of the pulling
force {\it vs} extension for a loading rate of $2.5\times 10^{-3}$m/s
is shown in Figure~1(A) inset (a) of the main text, for the residues centered at the midpoints
of the first candidate epitopes/anti-epitope (see Methods
). The first (weak stability) epitope
contains residues $5-15$ so the C$_\alpha$ atom of residue $10$ is
taken as a tethering point. The first anti-epitope predicted to be
thermo-mechanically stable consists of residues $16-20$ for which
residue $17$ is chosen as representative (see Methods).

The forces fluctuate stochastically, however the work to pull to a
distance $x$, being the integral of the force $W(x) = \int_0^x \!
F(x') dx'$ results in a smooth curve (Fig.~1(A)). The work
generally does not have a slope of zero as $x\rightarrow 0$ on the length
scale of $\sim 1 $\AA, because of an initial small-distance nonlinear
response corresponding to a steep rise in force within
$\sim 0.1$ \AA. That is, a force response function that appeared to
converge to a non-zero force as $x\rightarrow 0$ would correspond to a
work function with linear behavior as $x\rightarrow 0$.

We interpret the initial steep rise in force as being due the
collective effect of numerous strong bonds which seek to preserve the
native structure. As distance is increased, the number of restoring
interactions, and/or the magnitude of these interactions, is
decreased. Thus the effective modulus of the system as calculated by
$2 W(x)/x^2$ is distance
dependent, and softens with increasing distance (see Methods). A plot of the
effective modulus for short distances $< 1$\AA~ is given in inset (b) of Figure~1(A).

Previous measurements of force vs. extension or force vs. time have
shown that the force converges to non-zero values at short distances
or times. This is the case for ligand binding
simulations~\cite{ColizziF10} where the force converged to $\sim 50-100$~pN
for the shortest times, and in protein unfolding
simulations~\cite{LuH98,LuH99} where the force converged to $\sim
400-700$~pN at the shortest distances. These
observations are consistent with the steep initial rises in the force and corresponding
distance-dependent moduli that we have resolved in the present study.

It was also observed that pulling on a given residue resulted in large
fluctuations in remote regions of the protein. For example pulling on
residue $10$ disordered $\alpha$-helix 2 containing residues
133-138, and pulling on
residue $17$ disordered $\alpha$-helix 1 containing residues
55-61 (Figure~1(A) inset figures).
Local mechanical strain, at least by pulling a residue,
induces a non-trivial stress profile that results in induced disorder
at remote regions in the protein. Such induced disorder may be a
key ingredient in the propagation of misfolded SOD1 conformations in
ALS, as well as other misfolding diseases propagated by template
directed misfolding.

\subsection{The origin of large stiffness moduli at very short (sub-Angstrom)
  distances is likely due to side chain docking}
\label{sectsidechaindock}

What is the origin of this highly local mechanical rigidity that
gives rise to steep initial increases in force?
From our pulling simulations, it was observed that the forces required
to extend
well-structured parts of the protein were much larger than the forces
required for parts of the protein that were poorly structured or
disordered.
For example,
in the range of extensions from $\approx 0.1-0.2$\AA, the force on
residue 17 in $\beta$-strand 2 of SOD1 was $\approx 83 pN$, and the force
on residue 10 in turn 1 was $\approx 67 pN$, while the force on
residue 60 in the disordered Zn-binding loop of Zn-depleted
SOD1 was $\approx
52 pN$. As an example of a residue that should lack any side-chain
docking, the force on residue 133 at the disordered, non-native
C-terminus of the C-terminal truncation mutant E,E G127X is $\approx 41 pN$.
We thus investigated the
phenomenon of short-range mechanical rigidity by calculating the components of the interaction energy as
a tethered residue was pulled.

Figure~\ref{figInteenergy}(a) depicts a schematic of the simulation protocol,
and Figure~\ref{figInteenergy}(b) shows the results.
From these potential energy calculations, we see that the initial steep rise
in force is due to the loss of short-range van der Waals and electrostatic interactions
during the course of unfolding. The decrease in interactions is mainly between
side chains (SCs) rather than backbone (Fig.~\ref{figInteenergy}(b)
panel J):
roughly $3/4$ of the change in energy
arises from SC-SC interactions.
This effect thus appears to be due to the many-body interactions
stabilizing native structures through SC-SC docking, likely formed in the latest
stages of folding.

\subsection{Utility of making cumulative distribution}
\label{seccumu}

In the manuscript, we have analysed the mechanical profiles mainly by constructing cumulative
distributions of the work profiles, rather than the more common probability distribution
measurements. The reason behind representing the work profiles in the form of cumulative
distributions is that this representation gives the best way to differentiate between work profiles
of two different variants of SOD1. The mechanical work profiles are too noisy to compare their
relative stability from the sequence-resolved work profiles. Histograms of two work
profiles also do not clearly differentiate between two different variants of SOD1 (see Figure~\ref{figdistwork}).
However, the cumulative distributions of the work profiles make them completely distinguishable and
help to easily identify the relative order of stability among the variants.

\subsection{The mechanical profiles of Cu,Zn (SS) WT and E,E (SH) WT SOD1 are
  different, and are independent
  of the starting PDB structure used to construct them}
\label{sectmechprofile}

We can take the value of the work needed to pull a particular residue
out to 5\AA~as a representation of the mechanical rigidity of that
residue. This value can then be scanned across the protein sequence to
obtain a mechanical profile or fingerprint for a particular SOD1 variant. Obtaining
a work value for a given residue is computationally intensive however,
so we take a subset of $48$ residues as a ``sparse sampling''
of the mechanical profile, in order to compare mechanical stability
between SOD1 variants. The specific residues chosen are given in
the Methods section. Mechanical profiles may be compared
between WT SOD1 and various modified SOD1 proteins, including mutant
SOD1 (see Figure~\ref{figa4v}), 
de-metallated SOD1, and disulfide-reduced SOD1.

\renewcommand{\thefigure}{S\arabic{figure}}
\renewcommand{\arraystretch}{1}
\begin{figure}[h]
\centerline{\includegraphics[width=0.4\textwidth]{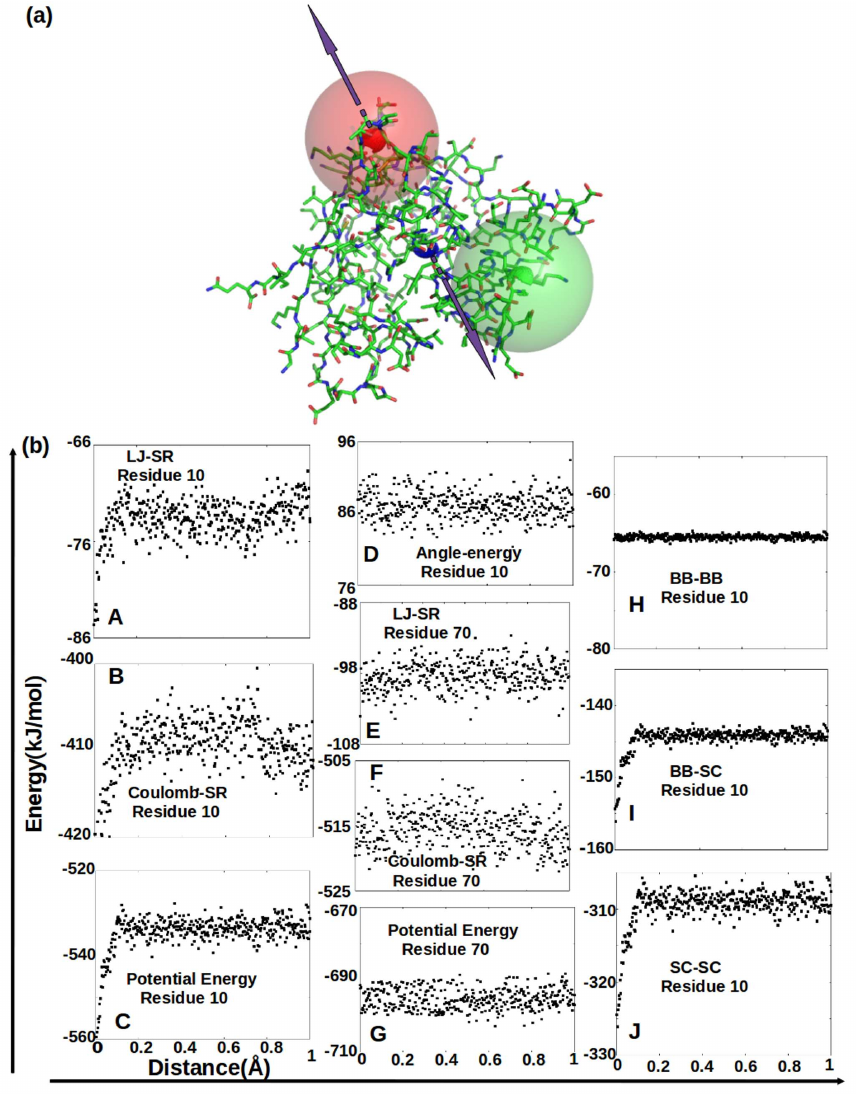}}
\caption{
(Panel (a))  Schematic representation of the method of calculation of
interaction energy terms shown in panel (b). We have calculated the interaction energy terms of the
atoms that are within $5$ \AA~ radius from the C$_\alpha$ atom of
residue $10$ and $70$ when residue $10$ is pulled -
the blue sphere indicates the C$_\alpha$ of residue $46$
which is the center of the protein, the small red sphere indicates the
C$_\alpha$ atom of residue $10$ which is the tethering point, and
small green sphere indicates the C$_\alpha$ atom of residue $70$. The
arrows show the direction of pulling.  The larger, semi-transparent
red and green spheres have radii of $5$ \AA, and enclose the atoms
within $5$ \AA~ from the respective C$_\alpha$
atoms of the protein.
(Panel (b)) Various terms in the potential energy as a function of
distance, when residue $10$ is pulled. The energies for residue $70$
are investigated as a control. Analyzing individual terms in the potential energy
elucidates the reason behind the initial sub-angstrom steep rise in
the mechanical force.
Figures A, B, and C plot the rise in short-range van der Waals energy,
short-range component of the Coulomb energy, and total potential
energy as a function of distance. These show a concurrent rise on the
length scale of the sudden rise in the force in
Figure~1(A) inset (a), main text.
No such distance-sensitive change is seen for angle energies in
residue 10 (figure D), or for any energies of a control residue (70) far from the pulling site
(figures E,F,G).
Decomposing the potential energy terms into backbone-backbone (H),
backbone-sidechain (I), and sidechain-sidechain terms (J) shows
that BB-BB interactions play no role, and about $3/4$ of the total contribution arises from SC-SC
interactions, indicating SC docking plays a dominant role in small
RMSD mechanical stability.
}
\label{figInteenergy}
\end{figure}

\renewcommand{\thefigure}{S\arabic{figure}}
\renewcommand{\arraystretch}{1}
\begin{figure}[h]
\centerline{\includegraphics[width=0.4\textwidth]{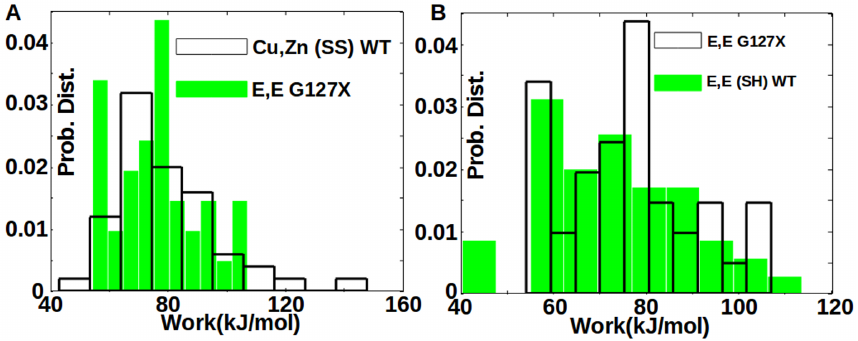}}
\caption{Panel A: Probability distribution of work values of Cu,Zn (SS) WT and
E,E G127X. Panel B: Probability distribution of work values of E,E G127X
and E,E (SH) WT SOD1.}
\label{figdistwork}
\end{figure}

We first ensured that the mechanical profile obtained for a given SOD1
variant was independent of the initial conditions used in the
simulations, in particular for SOD1 variants that currently have no PDB
structure such as E,E (SH) WT SOD1.  Equilibrated structures
were generated as described in the Methods
section below, and used as
initial conditions for pulling simulations to generate mechanical
profiles.
Inset A of  Figure~\ref{figStructinde} 
shows a plot of the mechanical scan for
Cu,Zn (SS) WT SOD1, and the main panel of Figure~\ref{figStructinde} shows the
mechanical scan for E,E (SH) WT SOD1. 

\renewcommand{\thefigure}{S\arabic{figure}}
\renewcommand{\arraystretch}{1}
\begin{figure}[h]
\centerline{\includegraphics[width=0.4\textwidth]{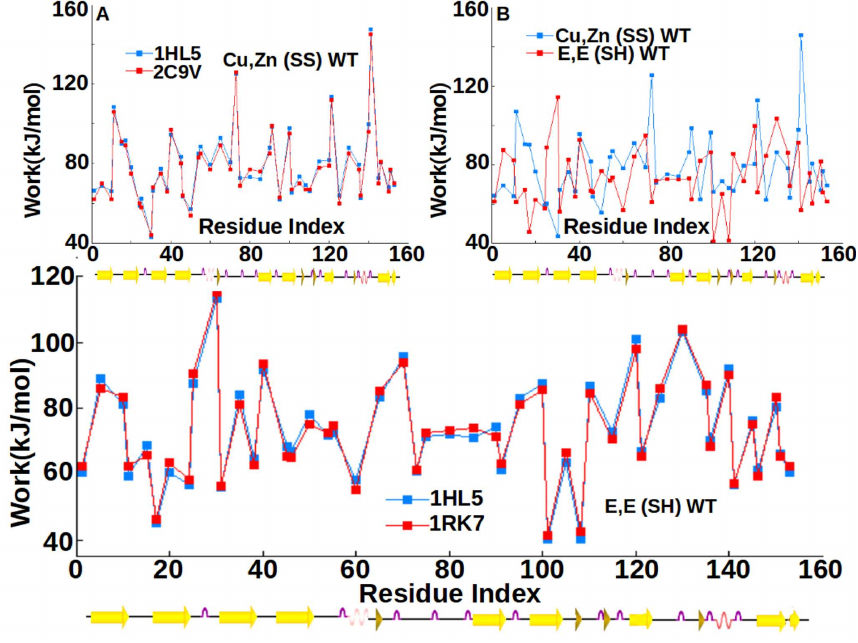}}
\caption{
The mechanical work profile is independent of the crystal/NMR
structure used to generate the initial ensemble in a pulling
simulation.
The main panel shows the mechanical profile
that results from 2 different constructions of the initial
ensemble of E,E (SH) WT SOD1. In one construction we start from the
solution structure of E,E (SS) WT SOD1, reduce the disulfide bond, and
equilibrate the system before starting simulations. In another
construction we start from the crystal structure of Cu,Zn (SS) WT SOD1, remove
the metals and reduce the SS-bond, and then equilibrate.
(Inset A) The mechanical profile obtained from 2 different
crystal structures of Cu,Zn (SS) WT SOD1 (1HL5 and 2C9V), equilibrated for $20$ns and then simulated as
described in the Methods. Both models give the same mechanical profile to within about $2.7$~kJ/mol.
(Inset B) Work profiles for Cu,Zn (SS) WT and E,E (SH) WT SOD1 are seen to
be significantly different, with E,E (SH) WT SOD1 generally having weakened
mechanical susceptibility in various regions, but occasionally showing
stiffer response in some locations.
}
\label{figStructinde}
\end{figure}

In each case, the protein
was constructed from two different initial models of the protein
structure, using two different PDB structures as starting points.
We found that the mechanical profile of a particular SOD1 variant was
nearly independent of how that variant was constructed, reinforcing the
reliability of the mechanical scan. In inset A of Figure~\ref{figStructinde}, mechanical work
profiles correspond to crystal
structures 1HL5~\cite{StrangeRW03} and 2C9V~\cite{StrangeRW06} of Cu,Zn (SS) WT
SOD1. In the main panel of Figure~\ref{figStructinde}, mechanical work profiles correspond to E,E (SH) WT SOD1,
obtained by modifying either the Cu,Zn (SS) WT
crystal structure 1HL5, or the NMR structure 1RK7 of
E,E (SS) WT SOD1~\cite{BanciL03}.
We note that these PDB
structures are equilibrated for $20$ ns before any simulation
measurements are taken.
The mean error between SOD1 variants, as determined by the z-test
described in the Methods section, is $2.7$ kJ/mol. On the
other hand, the standard deviation
of all 48 of the work values themselves for a given variant
(e.g. 1HL5) is $18.3$ kJ/mol, which is a
factor of about $6.8$ larger than the error. The mean error between
variants used to construct the same initial condition indicates the level of accuracy
of the simulations, so that differences in work profiles (e.g. between
mutant and WT) must generally be
larger than this mean error to be significant. Inset B to
Figure~\ref{figStructinde} plots the mechanical profiles of Cu,Zn (SS) WT SOD1 and
E,E (SH) WT SOD1. They are seen to be
significantly different; in particular the combination of metal depletion and
disulfide reduction reduces the overall mechanical stability of several
regions of SOD1. We analyze this in more detail in the manuscript.

\subsection{An implicit solvent model is sufficiently accurate to
  obtain the mechanical profile}
\label{sectimplicitexplicit}

To test the accuracy of the generalized Born surface area (GBSA) implicit solvent model in determining
the mechanical profile, we have performed pulling simulations on SOD1 in explicit
solvent, where waters interact through the SPC force field.

\renewcommand{\thefigure}{S\arabic{figure}}
\renewcommand{\arraystretch}{1}
\begin{figure}[h]
\centerline{\includegraphics[width=0.4\textwidth]{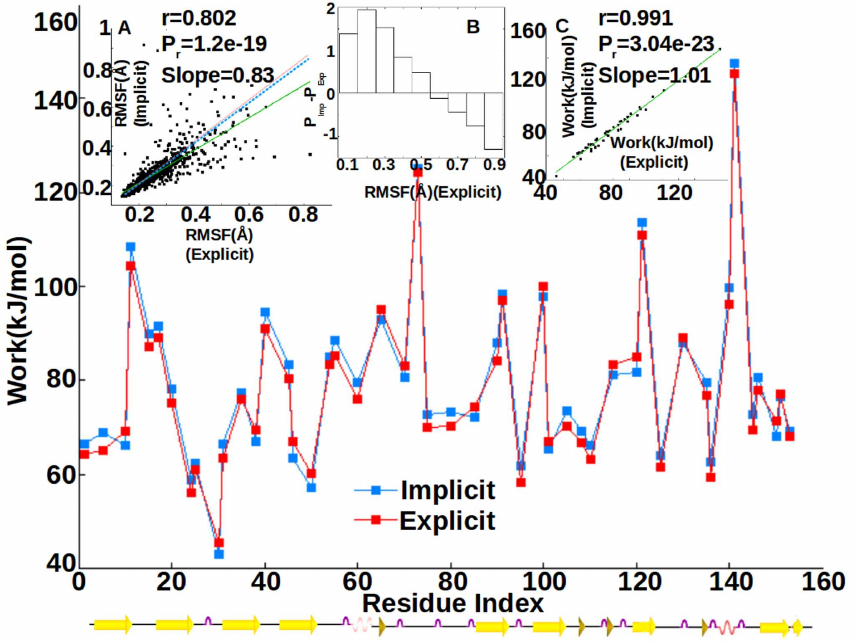}}
\caption{
The main panel shows that the work profiles agree between implicit
and explicit solvent simulations.
(Inset A) Scatter plot of the root mean square fluctuation
(RMSF) for heavy atoms in the protein
in both implicit and explicit solvents;
$r= 0.802$, $P_r=1.2\mbox{e-19}$. Green solid line is the best fit
line with a slope of $0.83$. Blue dashed line is the line of slope
unity, with $y=x$. Red dashed line is the median-fit line with an equal number of
data points above and below it, and has a sloe of $1.08$.
(Inset B)
Difference in the distributions of RMSF between the implicit and
explicit solvent models. This shows an enhancement of small RMSF
values and suppression of large RMSF values for the implicit solvent
model.
(Inset C) Scatter plot of work values obtained from implicit solvent {\it
  vs} explicit solvent models; these models correlate with $r=0.991$.
}
\label{figImplExplCorr}
\end{figure}

Inset A of Figure~\ref{figImplExplCorr} plots the root mean squared
fluctuations of heavy atoms, obtained from  20ns simulations for both
implicit and explicit solvent models. The two systems have comparable
thermal fluctuations, though the best fit line (green line in inset
A of Fig.~\ref{figImplExplCorr}) has a slope less than
unity, indicating somewhat larger fluctuations in the explicit solvent
model. Interestingly however,
the number of data points above and below the best fit line
are 624 and 455 respectively, indicating that there are non-Gaussian
fluctuations and outliers between the two models.

This skew in the data may be investigated through the distributions of
RMSF, for both the
implicit and explicit solvent models. These distributions are
different. The difference in the distributions,
$P_{Imp} - P_{Exp}$,  as a function of RMSF, is plotted in inset B
of Figure~\ref{figImplExplCorr}.
This shows
that the implicit solvent model overestimates small
fluctuations, and underestimates large fluctuations, as compared to
the explicit solvent model.

We can investigate what slope line would give equal numbers of data
points above and below it, as an additional measure of the validity of
the implicit solvent model. By this measure the implicit solvent model
agrees much better with the explicit solvent data: the median fit line
with equal numbers of data above and below it has a slope of nearly
unity (slope=1.08).

The imperfect correlation between implicit and explicit solvent
fluctuations prompts a comparison
of the work values in implicit and explicit solvent. A scatter plot of
work values to pull the same residues to 5~\AA~ in implicit and explicit solvent models is shown in inset
C of Figure~\ref{figImplExplCorr}. Interestingly, here we see a much stronger
correlation for the values of mechanical work. The mechanical work
values result from a significant non-equilibrium perturbation compared
to the local fluctuations in the native basin, the latter of which are apparently more
sensitive to solvent conditions.
A mechanical scan of
$48$ residues is shown in the main panel of
Figure~\ref{figImplExplCorr}. Here the implicit and explicit solvent
models show good agreement: the standard deviation of the difference
in work profiles is about $\approx 2.5$~kJ/mol which is less than
the mean error of $\approx 2.7$~kJ/mol obtained from using
different crystal structures to set up the same initial
conditions. One caveat is that the implicit solvent work values tend to be slightly
higher than those of the explicit solvent: the mean of $\Delta W$ is
about $1.1$~kJ/mol, so that a z-test indicates the data $\Delta W$
arise from a gaussian distribution of mean zero only when the standard
deviation of the gaussian distribution is $4$~kJ/mol or larger.
Overall, the data indicate that the implicit solvent model yields
mechanical profiles that are as reliable as those obtained from much more time-intensive explicit
SPC solvent simulations, but perhaps with modestly larger values
($\approx 1$~kJ/mol) of work.

\subsection{A G\={o} model does not
  adequately capture the mechanical profile to sufficient accuracy}
\label{sectGoImplicit}

Since the implicit-solvent model captured the mechanical profile to
good accuracy, we pursued a further step in simplifying the energy
function, to see if
a G\={o} model~\cite{GoN78}
would succeed in reproducing the mechanical profile. The
G\={o} model recipe~\cite{WhitfordP09} (see Methods) takes heavy atoms
within $2.5$ \AA,
and applies native contacts to them with an LJ-like 6-12
potential. The G\={o} recipe also attributes energy to native-like
dihedral angles. The overall energy scale of all interactions is given by $1$~kJ/mol
times the number of atoms in the system.
This recipe is intended to
approximately account for all native stabilizing interactions as well as solvation free
energy.

Figure~\ref{figAllatomGoCorr} plots the work profiles of Cu,Zn (SS) WT SOD1,
for both an all-atom implicit-solvent model and an all-atom G\={o}
model.
Perhaps surprisingly, the default energy scale in the G\={o} model,
$1$~kJ/mol times the number of atoms, captures the overall energy
scale of the work profile quite well: both
energy functions resulted in variation of the work from about
$40$~kJ/mol to about $120$~kJ/mol.
However, from a blind comparison of the cumulative distributions for the implicit
solvent and G\={o} models, one would conclude they were different
proteins, so the distribution of work values is significantly
different. One can adjust the overall energy scale in the G\={o} model
to better capture the mechanical work distribution, but the optimal
value of the energy scale is not known {\it  a  priori}.
Moreover, the correlation between the implicit-solvent and G\={o}
models, $r=0.377$, is not strong (Fig.~\ref{figAllatomGoCorr} inset
A). Increasing the overall G\={o} energy scale by a factor of $1.1$ to improve the comparison of
the cumulative distributions (inset C of
Fig.~\ref{figAllatomGoCorr}) does not improve the correlation
between work values: $r=0.385$, $P=.007$.
The green line in inset A of Figure~\ref{figAllatomGoCorr}
indicates the best linear fit between the G\={o} and implicit solvent
models. The most significant outlier on the scatter plot is residue
141, a glycine
(circled data point in Fig.~\ref{figAllatomGoCorr} inset A),
which also can be seen to have the largest discrepancy in the
mechanical profile. It has the largest work in the implicit-solvent
model, and one of the smallest in the G\={o} model.
Excluding this residue increases the correlation between the two
models to $0.621$ (blue line in inset A). Why is it so anomalous? The residue resides in the
so called electrostatic loop, which is enriched in charged and polar
residues, and coulombic energies are not explicitly treated
in the G\={o} model, which only contains 6-12 van der Waals-like
interactions.

\renewcommand{\thefigure}{S\arabic{figure}}
\renewcommand{\arraystretch}{1}
\begin{figure}[h]
\centerline{\includegraphics[width=0.4\textwidth]{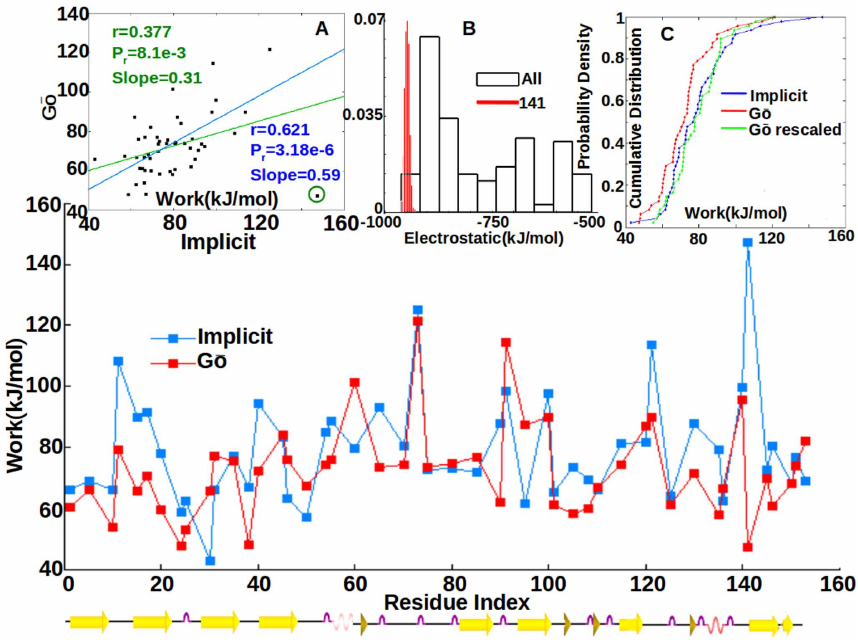}}
\caption{
Work profiles of Cu,Zn (SS) WT SOD1 obtained from an implicit
solvent model, and a G\={o} model. Both models are all-atom.
(Inset A) Scatter plot of the work values obtained from both models.
The green line is the best fit line to the data, which has a
correlation that is weak ($r=0.377$)
but statistically significant ($P=0.008$). Note that the energy scales range from
about $40$kJ/mol to $120$kJ/mol for both models.
Omitting one amino acid in the electrostatic loop, residue G$141$
(green circled data point in the lower right of panel A),
increases the correlation to $0.62$ and the significance to
3e-6 (blue line). Either with or without residue $141$, the slope of the line is
less than unity however, indicating that stabilizing energetics are missing in the G\={o}
model.
(Inset B)  Distribution of the electrostatic potential energy within a
sphere of radius 5\AA~ centered at the C$\alpha$ atom, for all residues in the monomeric
protein. Residue $141$ has one of the largest contributions of
electrostatic energy, which explains why its work value in the
implicit solvent model was much higher than that in the G\={o} model,
which does not explicitly account for electrostatics.
(Inset C) Cumulative distributions of the work values obtained from the
implicit solvent (blue) and G\={o} (red) models.
The mean work difference in the cumulative distributions between the
two models is $\approx 7$ kJ/mol.
The green cumulative distribution in inset C corresponds to a G\={o}
model that has been reweighted to have contact and dihedral energies
that are $1.1\times$ as strong. This shifts the work distribution to
larger values, but the values themselves still do not correlate well with those
in the implicit solvent model: $r=0.385$, $P=0.007$.
}
\label{figAllatomGoCorr}
\end{figure}

We thus investigated the electrostatic component of the
energy within a sphere of radius 5\AA, centered at the C$\alpha$
atom for every residue, to construct the histogram in inset B of
Figure~\ref{figAllatomGoCorr}. The energies plotted are the mean
values of the energy from an equilibrium simulation at 300K. The
histogram of electrostatic energy for glycine 141 is also plotted.
Indeed, residue 141 has one of the largest electrostatic contributions
to its energy. This is impressive because of the small size, apolarity,
and neutrality of the residue. Electrostatic contributions to protein stability, for
example due to ion pairs or partial charges in either close proximity
or in low dielectric environment, may be poorly accounted for in
G\={o} models.

\subsection{The mechanical profile accurately reflects the free energy profile
  of the protein}
\label{whamfreeE}

The work to pull a given C$\alpha$ atom to 5\AA~is a non-equilibrium
measurement of mechanical stiffness, and one can ask whether it accurately represents the
thermodynamic stability of that region. To address this question, we
obtained the free energy to separate each of the $48$ C$\alpha$ atoms
used in the mechanical pulling simulations by 5\AA. The procedure for
obtaining the free energy is described in the Methods section.
Figure~\ref{figWorkFreeCorr} plots the work values for all $48$ residues used in the pulling
simulations of Cu,Zn (SS) WT SOD1, {\it vs.} the free energy values for the
corresponding residues as obtained from the weighted histogram
analysis method (WHAM). The correlation
coefficient is $0.96$, indicating that the relative mechanical
rigidity can be used to predict the relative thermodynamic stability.

\renewcommand{\thefigure}{S\arabic{figure}}
\renewcommand{\arraystretch}{1}
\begin{figure}[h]
\centerline{\includegraphics[width=0.4\textwidth]{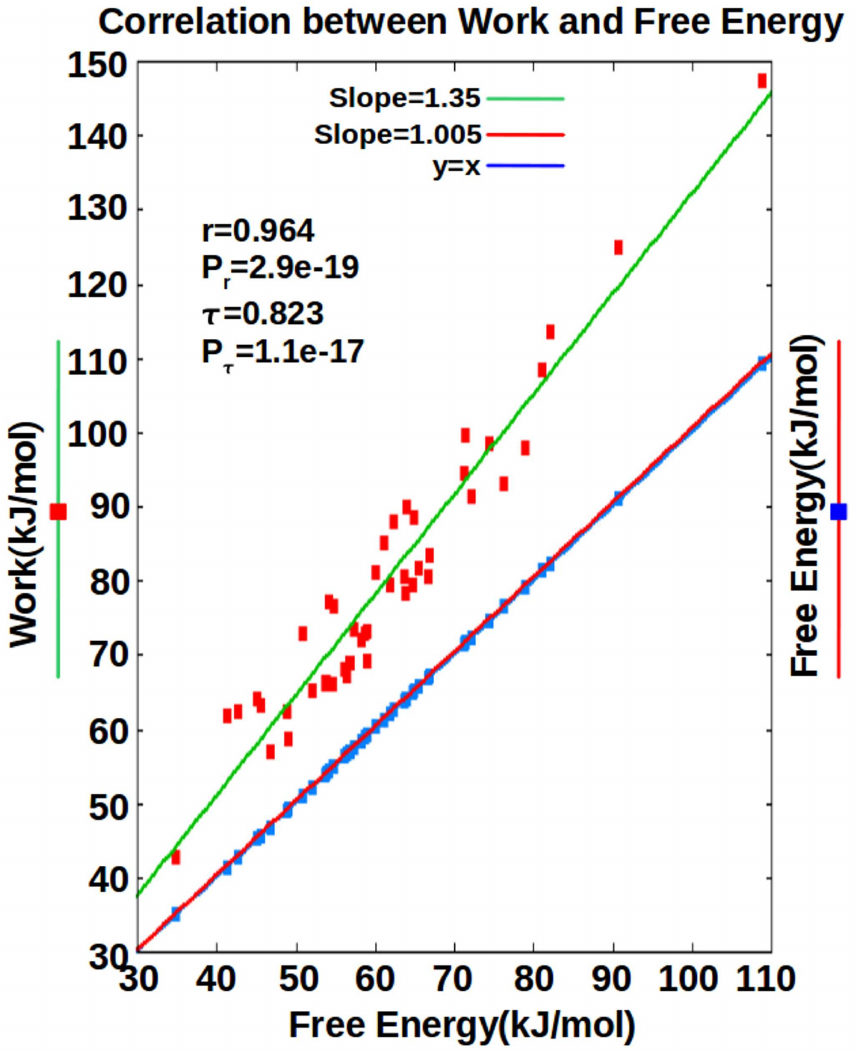}}
\caption{
(Red squares) Work to pull a given residue to 5\AA~{\it vs} the free
energy change for a fluctuation to separate that residue by 5\AA, for
Cu,Zn (SS) WT SOD1. Free
energies on the abscissa are obtained from umbrella sampling and the weighted
histogram analysis method (WHAM). Both work values and free energies,
for the red squares, are obtained using a slow pulling speed of
$2.5\times 10^{-3}$m/s.
All $48$ residues used in the
pulling simulations are shown. The correlation between work and free
energy values is very strong, $r=0.96$.
For the pulling rates used in our study, the work values are about
$1.3\times$ as large as the free energies, and the slope of the
best fit line is $1.35$.
Blue squares:
WHAM-derived free energy changes, as obtained from pulling
simulations at two different pulling rates:
The abscissa values give the free energy for a pulling rate of
$2.5\times 10^{-3}$m/s as above,
the ordinate values have a pulling rate of $10$ m/s.
In addition to a near perfect correlation, the slope defined by the
best fit line to the
two sets of data is nearly unity (purple line), and the data is well-fit by the line
$y=x$ (cyan line) indicating that the free energy
values are independent of
the pulling speed used to obtain the initial data, and thus
have been reliably determined.
}
\label{figWorkFreeCorr}
\end{figure}

The work values are higher than the free energy values however, by a
factor of about $\left< W/F\right> \approx 1.3$.
The average mechanical work, as a non-equilibrium
measurement, always exceeds the free energy change that would be due
to rare equilibrium fluctuations.

Since a faster pulling rate results in more deformation of the
protein, different pulling
rates can in principle result in initial conditions that, after
umbrella sampling and WHAM, give different free energy profiles.
We checked this by performing WHAM calculations at two different
pulling speeds, $2.5\times 10^{-3}$m/s and $10$ m/s. The faster
pulling speed results in more deformed protein structures that were
used as initial conditions, however, each initial condition is always
equilibrated for 10 ns in an umbrella potential as described in
the Methods, which should remove most or all initial deformation effects.
The free energy values
thus obtained did not depend on the initial pulling speed used to
generate the initial conditions for the WHAM protocol: they are within a factor of $1.005$.
We thus used relatively fast pulling speeds of 10 m/s to obtain
initial conditions used to calculate free
energies from the WHAM method.

\subsection{The mechanical profile obtained from about 40 residues captures
  the distribution of work values for a given SOD1 variant to
  sufficient accuracy}
\label{workdistrib}

Inset A of Figure~\ref{figConvergence} shows the mechanical work
profile to pull residues to $5$\AA, for every residue between the
N-terminus and residue $40$. Residues from the original set of 48 are
shown in red, others in blue. The values deviate significantly
residue to residue, with a correlation length less than the putative
value of $\sim5$ amino acids corresponding to the original data
set. By calculating the
residue-residue correlation function of the work $\left< W_i \cdot
    W_{i+n} \right> $ and fitting to  $\exp (-n/\ell_p )$,
the sequence correlation length is found to be $\approx 2.83$.
The work values found from the mechanical scan using the original sampling of $48$ residues
should thus not be interpreted as consensus values for the
corresponding regions.

\renewcommand{\thefigure}{S\arabic{figure}}
\renewcommand{\arraystretch}{1}
\begin{figure}[h]
\centerline{\includegraphics[width=0.4\textwidth]{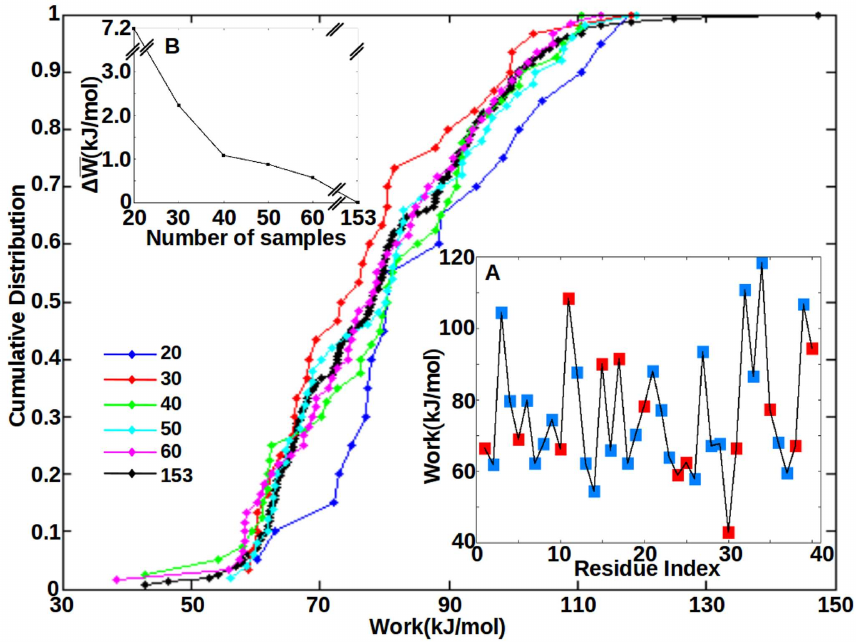}}
\caption{
(Inset A) Mechanical work
profile to pull residues to $5$\AA, for every amino acid between the
N-terminus and residue $40$. Residues from the original set of $48$,
discussed in the Methods subsection on residues used for mechanical scans, are
shown in red, others in blue. Work values deviate significantly
residue to residue, with a sequence correlation length $\ell_p \approx
2.83$, as defined through $\left< W_i \cdot
    W_{i+n} \right>  \propto \exp (-n/\ell_p )$.
The work values found from the mechanical scan using the original set of $48$ residues
should thus not be interpreted as consensus values for the
corresponding regions.
(Main panel) Cumulative
distributions of the work values for Cu,Zn (SS) WT SOD1, constructed simply
by rank ordering the work values and plotting the fraction of the
total number vs the work. Each cumulative distribution
corresponds to a specific number of data points (work values) as given in the
legend, which are randomly selected from the
total set of $153$ data points.
As the number of data points is increased, the cumulative distribution
converges to that for the full data set. The average deviation of work
values is plotted in inset B. By 40 data points, the cumulative
distribution has converged to within $\approx 1.08$ kJ/mol of the
full $153$-data point distribution. By 50 data points, it has converged
to within $\approx 0.88$ kJ/mol of the full distribution.
}
\label{figConvergence}
\end{figure}

One can ask if the mechanical scan still has utility then.
The main panel of Figure~\ref{figConvergence} shows a cumulative
distribution of the work values for Cu,Zn (SS) WT SOD1, which is constructed simply
by rank ordering the work values and plotting the number vs the work,
subsequently normalizing to unity. Several curves are shown: each curve
corresponds to a given number of data points randomly selected from the
total set of $153$ data points.  As the number of data
points is increased, the distribution converges to that containing the
full data set. Inset B of Figure~\ref{figConvergence} plots the
mean deviation in work values between cumulative distributions. After
about $40$ data points, the cumulative distribution converges to within 
about $1.08$ kJ/mol 
of the full distribution using $153$ data points.
This deviation is smaller than
the deviation in work values obtained from different starting PDB
structures (see above subsection ``The mechanical profiles of Cu,Zn
(SS) WT ..'' and Fig.~\ref{figStructinde}). That is, if one constructs 
cumulative distributions from the work values in
Figure~\ref{figStructinde}, the mean work difference between
cumulative distributions is $\approx 1.2$kJ/mol.
The data set used for our analysis, given in the Methods
section, contains $48$ residues, and has a mean deviation
from the $153$-residue cumulative distribution of about $0.9$
kJ/mol.

We found 
that many discrepancies of SOD1 variants
from WT were difficult to disentangle from the work profile, but often
emerged naturally from the cumulative distribution.
Mechanical scans were thus used to construct the cumulative
distributions, which then allowed us to distinguish
stabilizing energetics in various forms
of mutant SOD1. For example, we find below
that the cumulative distribution is different for differently metallated
variants of SOD1 (Figure~\ref{figmetalss}).

\subsection{Equilibrium dynamical results}
\label{workdistrib}

\renewcommand{\thefigure}{S\arabic{figure}}
\renewcommand{\arraystretch}{1}
\begin{figure}[h]
\centerline{\includegraphics[width=0.5\textwidth]{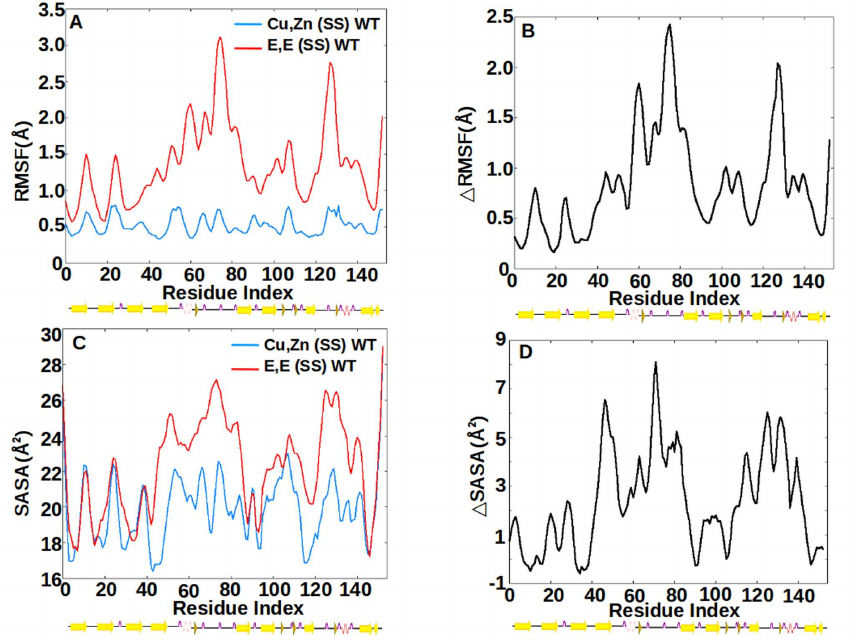}}
\caption{
Conformational disorder in the absence of metals, as obtained from
equilibrium simulations, is largest in
Zn-binding and electrostatic loops.
Comparison of root mean squared fluctuations (RMSF) (A,B) and solvent
accessible surface area of backbone amides (SASA${}_N$) (C,D), for
Cu,Zn (SS) WT and E,E (SS) WT SOD1. Results
are averaged over a 20 ns equilibrium simulation trajectory.
Panels A and C plot the respective
quantities for both proteins, and panels B and D plot the
differences between E,E (SS) WT and Cu,Zn (SS) WT proteins, i.e. the change in RMSF or
SASA${}_N$ upon loss of metals. Both quantities indicate substantial
increase in dynamic disorder and loss of native structure in loops IV
and VII, and to a lesser extent loop VI, upon loss of metals.
}
\label{figholoapo}
\end{figure}

Removing metals from WT protein results in substantially increased
dynamical fluctuations (RMSF) in loops IV (ZBL) and VII (ESL), and to
some extent loop VI (Figure~\ref{figholoapo}). The RMSF of Cu,Zn(SS) WT and E,E(SS) WT, along
with their difference, $\Delta$RMSF, are shown in
Figures~\ref{figholoapo}A,B. Loops IV and VII also show increased
solvent accessible surface area of backbone amide nitrogen
(SASA${}_N$) upon loss of metals (Fig.~\ref{figholoapo} panels C,D). 

The dynamic effects of fluctuations between E,E(SH) and G127X seen in
Figure~2C of the main text are not
recapitulated in backbone amide solvent exposure:  loop IV of E,E (SH) WT SOD1 is
not significantly more solvent exposed than loop IV in E,E G127X, and the
N terminus of E,E G127X from strand $\beta$7 onwards is more solvent
exposed than the corresponding residues in E,E (SH) WT SOD1
(Fig.~\ref{figsasa}).

The profile of potential energy can be obtained by
finding the potential energy within a sphere of radius 8\AA, centered
about a given $C\alpha$ atom, 
and then varying the $C\alpha$ index  
along the
sequence. Consistent with the increased mechanical stability of G127X with
respect to E,E(SH) WT (Fig.~1B), the potential energy profile of G127X
is generally more stabilized than that of E,E(SH) WT
(Fig.~\ref{figpot}). 

\renewcommand{\thefigure}{S\arabic{figure}}
\renewcommand{\arraystretch}{1}
\begin{figure}[h]
\centerline{\includegraphics[width=0.4\textwidth]{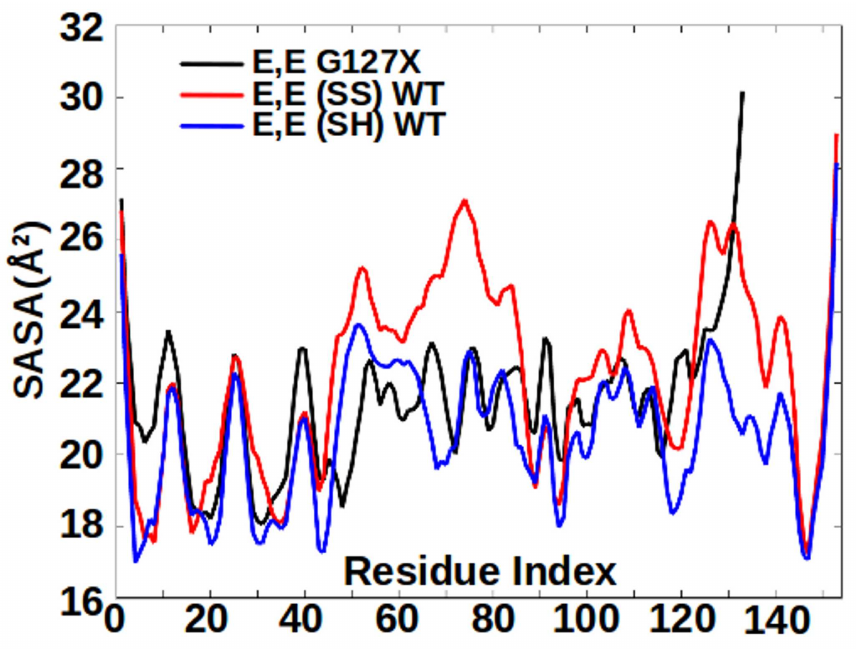}}
\caption{
SASA of backbone amide nitrogen atoms, from
equilibrium simulations of E,E G127X, E,E(SS) WT and E,E(SH) WT SOD1.
}
\label{figsasa}
\end{figure}

\renewcommand{\thefigure}{S\arabic{figure}}
\renewcommand{\arraystretch}{1}
\begin{figure}[h]
\centerline{\includegraphics[width=0.4\textwidth]{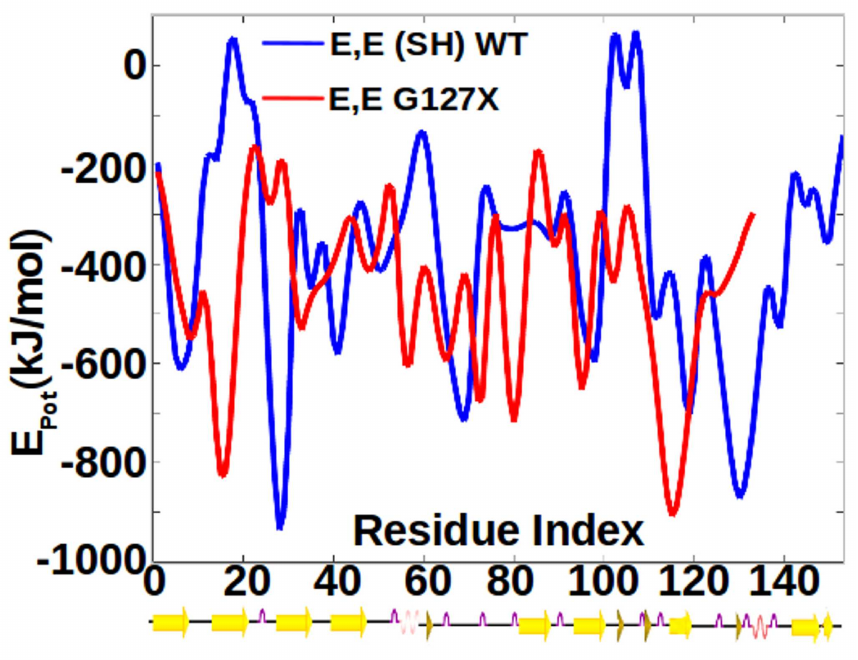}}
\caption{
Interaction potential energy between all atoms within an 8 \AA~ sphere of
the $C_\alpha$ atom of each residue, for E,E(SH) WT SOD1 and E,E G127X
SOD1.}
\label{figpot}
\end{figure}

\subsection{Comparison of metal binding free energy, thermodynamic
  stability, and mechanical stability}
\label{compmut}

The thermodynamic stability of the E,E(SS) monomer for several SOD1
mutants has been obtained previously by Oliveberg and 
colleagues~\cite{LindbergM2005}. 
The free energy difference between WT and mutant free energy of unfolding is
plotted in Figure~\ref{figthermo}. As well, the change in average
mechanical work values, from WT to mutant, is plotted in the same
figure. The work values have been normalized by a factor of 1.35, the slope
of Figure~\ref{figWorkFreeCorr}, to convert work values to effective free
energies.
Finally, the difference in metal binding free energy,
WT minus mutant,
for both Cu and Zn, is plotted in the same figure. All free energy
changes are positive and between 1-25 kJ/mol. 
However, there is no significant
correlation between any of these quantities, with the exception of the
binding free energies of Cu and Zn
($r=0.91,p=6.6\mbox{e-4}$).

\subsection{Cumulative distributions of mechanical work for an ALS-associated mutant}
\label{workmut}

Figure~\ref{figa4v} shows the cumulative distributions of mechanical
work for both WT SOD1 and A4V, for several different post-translatinal
modification (PTM) states. The
Cu,Zn(SS) mechanical stabilities of WT and A4V are very similar, except for the
weaker regions. In contrast, the E,E(SS) protein is substantially
mechanically destabilized upon mutation, and the E,E(SH) protein is
even further mechanically destabilized upon mutation. Interestingly,
over much of the distribution E,E(SH) WT is stabilized with respect to
E,E(SS) WT. This stabilization is also observed by examining the
change in potential energy upon disulfide reduction~\cite{DasA12jmb}. 

\renewcommand{\thefigure}{S\arabic{figure}}
\renewcommand{\arraystretch}{1}
\begin{figure}[h]
\centerline{\includegraphics[width=0.4\textwidth]{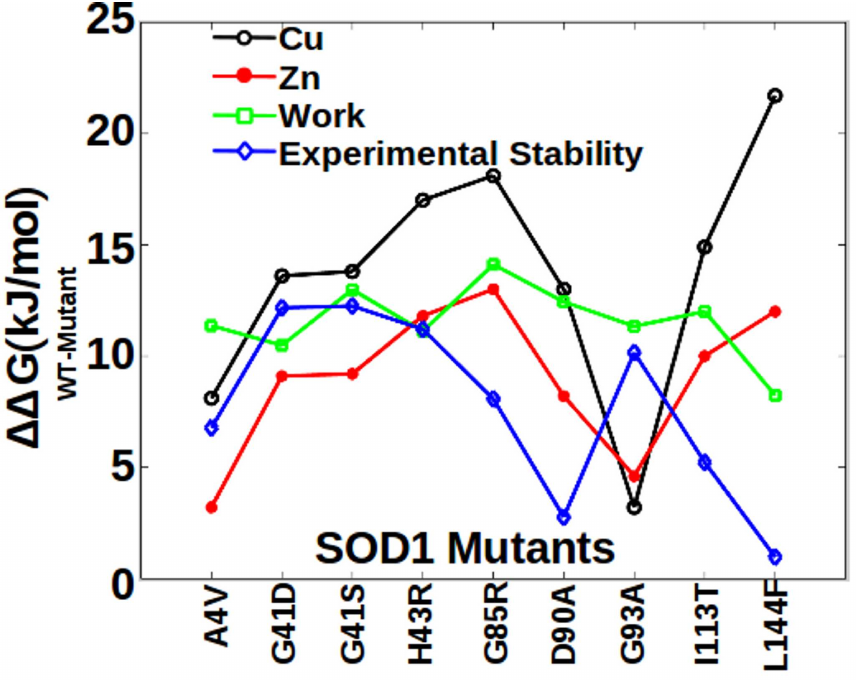}}
\caption{
(Black) Difference in free energy of Cu binding, WT minus
mutant, (Red) Difference in free energy of Zn binding, WT minus
mutant,  (Green) Difference in average mechanical work values, WT
minus mutant. The work values have been normalized by the slope
of Figure~\ref{figWorkFreeCorr} to convert them to effective free
energies. (Blue) Difference in experimental thermodynamic
free energy of unfolding~\cite{LindbergM2005}, WT minus
mutant.
}
\label{figthermo}
\end{figure}

\renewcommand{\thefigure}{S\arabic{figure}}
\renewcommand{\arraystretch}{1}
\begin{figure}[h]
\centerline{\includegraphics[width=0.4\textwidth]{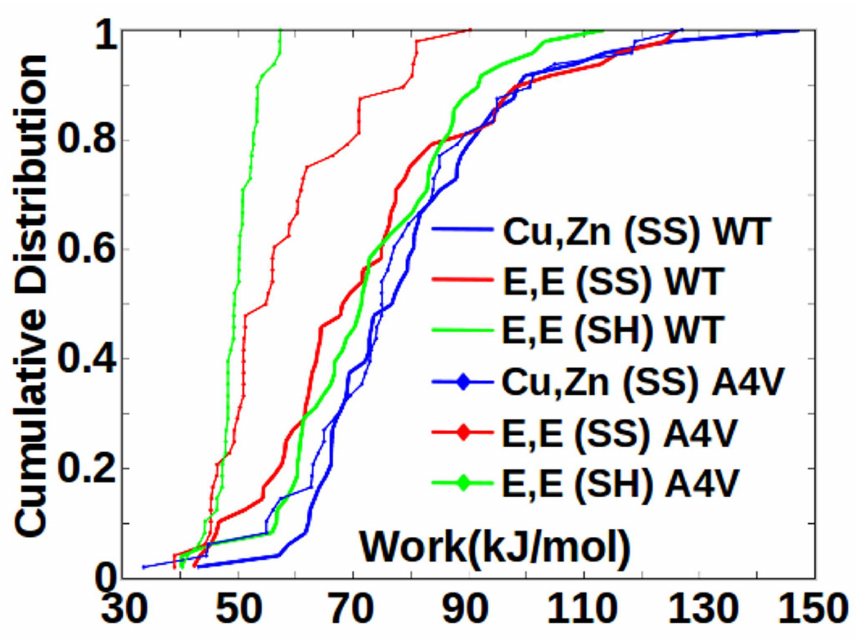}}
\caption{Cumulative distributions of work values for
  non-post-translationally modified (PTM) variants of
  WT SOD1, i.e. those missing metals and/or disulfide bond. Cumulative
  distributions of A4V SOD1 in various PTM states are also shown.}
\label{figa4v}
\end{figure}

The change in mechanical stability upon mutation for the E,E(SH) state
is in contrast to the change in mechanical stability upon truncation
for G127X (Figure 1B, main text). The mutations are quite different- one replaces an Alanine
with a Valine and the other more drastically removes 20 residues and mutates 6
C-terminal residues.

\subsection{Cumulative distributions of WT SOD1 PTM variants}
\label{workmut}

Figure~\ref{figmetalss} shows the cumulative distributions of
Cu,Zn(SS) SOD1 along with several
WT SOD1 PTM variants, including E,Zn(SS), Cu,E(SS), Cu,Zn(SH), and
E,E(SS) WT. Comparing the mechanical destabilization of the premature
variants with Cu,Zn(SS) SOD1, we see that E,E(SS) SOD1 is the
least mechanically stable. Comparing individual PTM states, the loss of Zn results in the
largest destabilization, followed by disulfide reduction, followed by
Cu depletion. 

\renewcommand{\thefigure}{S\arabic{figure}}
\renewcommand{\arraystretch}{1}
\begin{figure}[h]
\centerline{\includegraphics[width=0.4\textwidth]{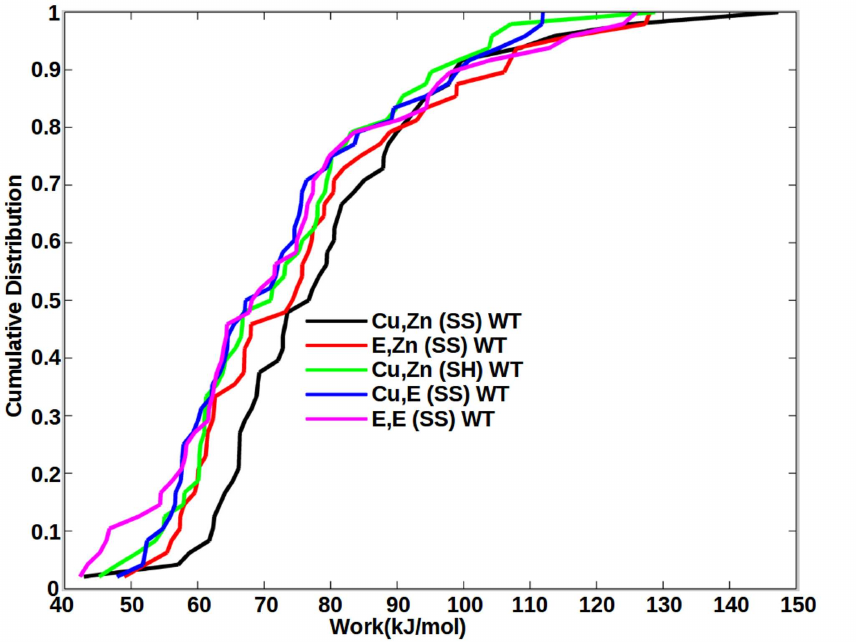}}
\caption{
Cumulative distribution of work values for various PTM states of WT SOD1,
as indicated in the legend.
}
\label{figmetalss}
\end{figure}

\subsection{ALS-associated mutants}
\label{workdistrib}
ALS-associated mutants considered in the main text were: 
A4V,
W32S, G37R, L38V, G41D. G41S, H43R, H46R, H46R/H48Q, T54R, D76Y,
H80R, G85R, D90A, G93A, G93C, I113T, G127X, D124V, D125H, S134N
and L144F.

\subsection{Robustness tests of the results}
\label{robust}

In the simulations, mechanical fingerprints were determined at
$T=300K$, and pulling speed $v=2.5$mm/s; both numbers are comparable
to those in AFM experimental assays. The pulling distance $\Delta x =
5$\AA was taken to represent a significant perturbation from the native
structure, but not so much so as to globally unfold the protein. 
To test the  robustness of our results to varying external conditions,
we have taken one of our conclusions, namely that E,E C57S 1-127 SOD1 is
mechanically stabilized with respect to E,E C57S/C146S 1-153 SOD1 (see
Figure~1D main text), and we have 
varied temperature, pulling speed, and pulling distance. Temperature
was increased to $310K$, pulling speed was altered to $v=1.5$mm/s and
$4$mm/s, and pulling distance was altered to $\Delta x=3,4,6$ and
$7$\AA. The results are summarized in Figure~\ref{figrobust}.


Panel A of Figure~\ref{figrobust} is at the conditions used throughout
our analysis, and reproduces Figure~1D of the main text. 
Increasing the temperature to biological temperatures ($T=310$K)
rather than lab temperatures preserves the conclusion and only slightly reduces the 
statistical significance. 
Decreasing or increasing the pulling speed to the values noted above tends to decrease or
increase all work values by $2$-$3$kJ/mol, but preserves the
conclusion and leaves the significance nearly unchanged. 
Panels E-H of Figure~\ref{figrobust} show increasing pulling distance
$\Delta x$ from $3$\AA to $7$\AA. The conclusions are preserved for
all distances, and the statistical significance of the conclusion increases with
increasing distance pulled. Taken together, the above tests indicate that
the conclusions in the manuscript are robust to varying external
conditions. 


\subsection{Work values projected on SOD1 structure}
\label{project}

A mechanical scan was performed on Cu,Zn(SS) WT SOD1 for every
$C\alpha$ atom, and the work values were then projected
on the native protein structure in Figure~\ref{figproject} panel A. As
mentioned in the text, the correlation length of the work values along
the sequence is about $3$, so the projection color-coded by work value
is discontinuous. Applying to the work values a simple tent-shaped smoothing function
that smoothes over 5 residues gives the plot of 
Figure~\ref{figproject} panel B. 
This shows that for example parts of loops IV and VII are mechanically
more rigid than some beta sheets, indicating significant stabilization
by the presence of the metals.

\renewcommand{\thefigure}{S\arabic{figure}}
\renewcommand{\arraystretch}{1}
\begin{figure}[h]
\centerline{\includegraphics[width=0.5\textwidth]{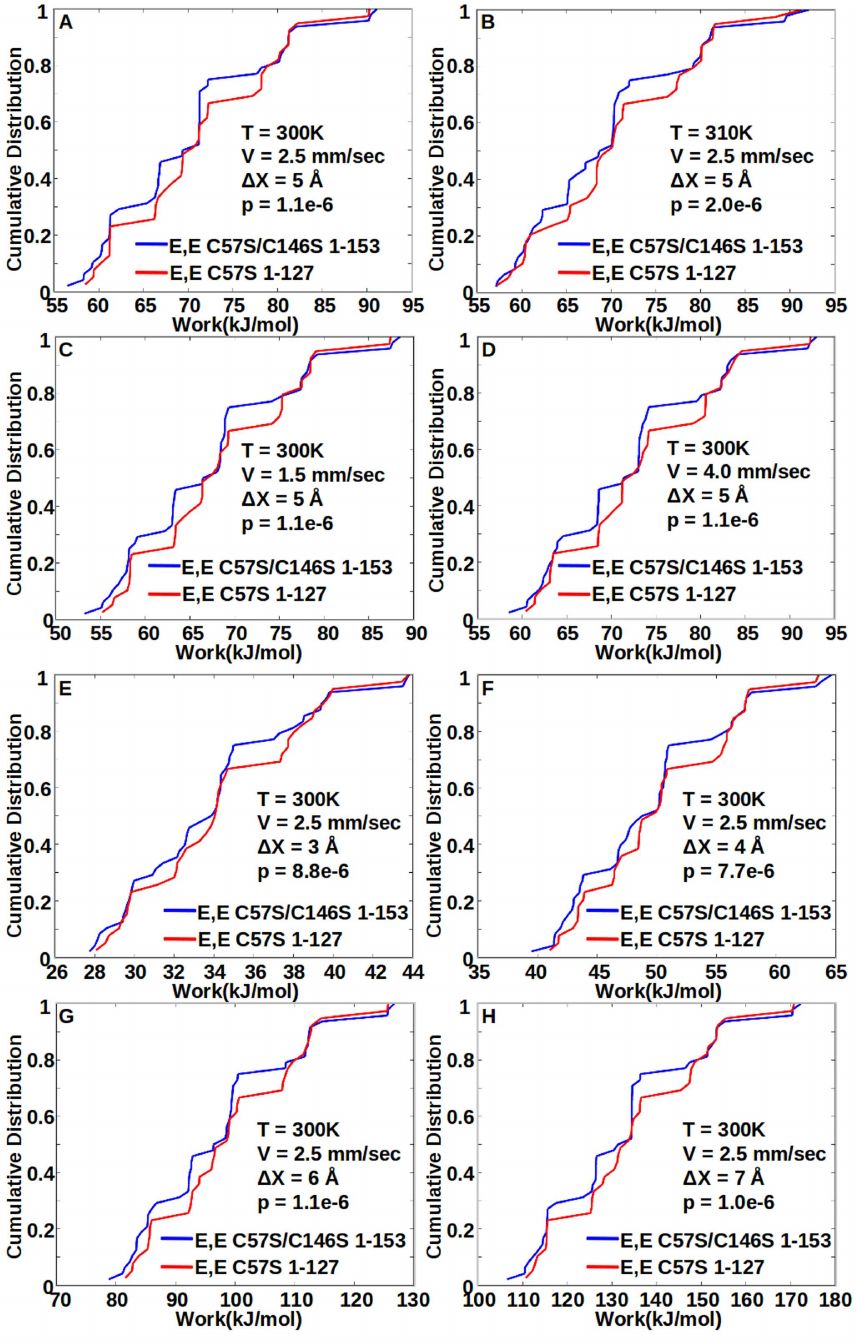}}
\caption{Cumulative distribution of mechanical profiles showing E,E C57S 1-127 SOD1 is
mechanically stabilized with respect to E,E C57S/C146S 1-153 SOD1 (see
Figure~1D main text). Temperature $T$, speed $v$, extension $\Delta
x$, and statistical significance of the result $p$ as defined in the
Methods below, are given for each panel.
(Panel A) Assay done at conditions generally used in the main text.
(Panel B): Same as panel A but at $T=310$K.
(Panels C,D): Same as panel A but at pulling speeds $v=1.5$mm/s and
$4$mm/s.
(Panels E-H): Same as panel A but at pulling distances 3,4,6 and 7
\AA. Statistical significance monotonically increases with pulling
distance.
}
\label{figrobust}
\end{figure}


\subsection{Frustratometer analysis of a non-ALS related alanine scan of SOD1}
\label{nonals}

Figure~\ref{figapohf} shows the results frustratometer analysis for
both Cu,Zn(SS) SOD1, and E,E(SS) SOD1. The mean number of highly
frustrated contacts for each variant is $2.2$ and $2.45$ respectively,
indicating that the E,E(SS) state is more frustrated by about 38
contacts. 
In the main text we used frustatometer analysis~\cite{FerreiroDU11} to
show that more frustration was present in the E,E(SS) state of SOD1
than the Cu,Zn(SS) state, by showing that upon mutation, more highly frustrated
contacts were introduced in the Cu,Zn(SS) state, but frustration was
removed by mutation in the E,E(SS) state.  A set of 22
ALS-associated mutations were taken. 
Here, we check that the conclusion is general- not restricted to
ALS-associated mutation- by examining the effects of 30 $X\rightarrow
A$ mutations. An alanine scan across the sequence is taken, wherein 
every 5th residue
in the SOD1 sequence is mutated to alanine if (1) it is not an alanine in
the WT sequence and (2) the mutation to alanine does not lead to an ALS-associated
mutant. When we encountered alanine in the WT sequence (residue 55, 60, 95, 140 and 145),
we took either the previous or the next residue for our analysis as
indicated below. As well, D90A is an ALS-associated mutation, so for
this case we chose residue 91, i.e. K91A. 

\renewcommand{\thefigure}{S\arabic{figure}}
\renewcommand{\arraystretch}{1}
\begin{figure}[h]
\centerline{\includegraphics[width=0.5\textwidth]{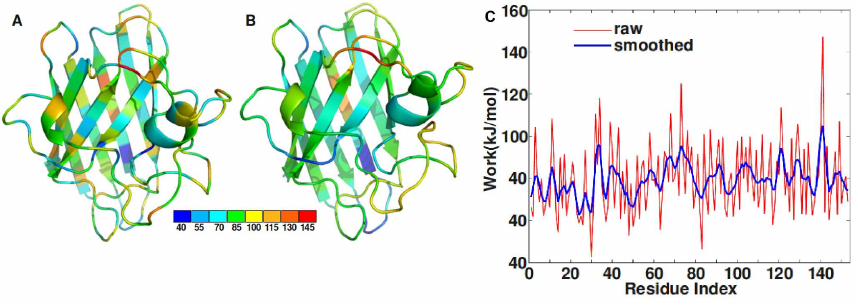}}
\caption{
(Panel A): Work values to pull all 153 residues to 5 \AA~for Cu,Zn(SS)
WT SOD1,
projected on the native SOD1 conformation - work values increases from
blue to red. The very low
correlation length of work values along the primary sequence is
evident from the discontinuity of the color scheme.
(Panel B): A smoothed work profile was generated from the all-residue work profile used in panel
A and projected on the SOD1 conformation.
(Panel C) The raw and smoothed work values as a function of
  sequence index.
}
\label{figproject}
\end{figure}

\renewcommand{\thefigure}{S\arabic{figure}}
\renewcommand{\arraystretch}{1}
\begin{figure}[h]
\centerline{\includegraphics[width=0.3\textwidth]{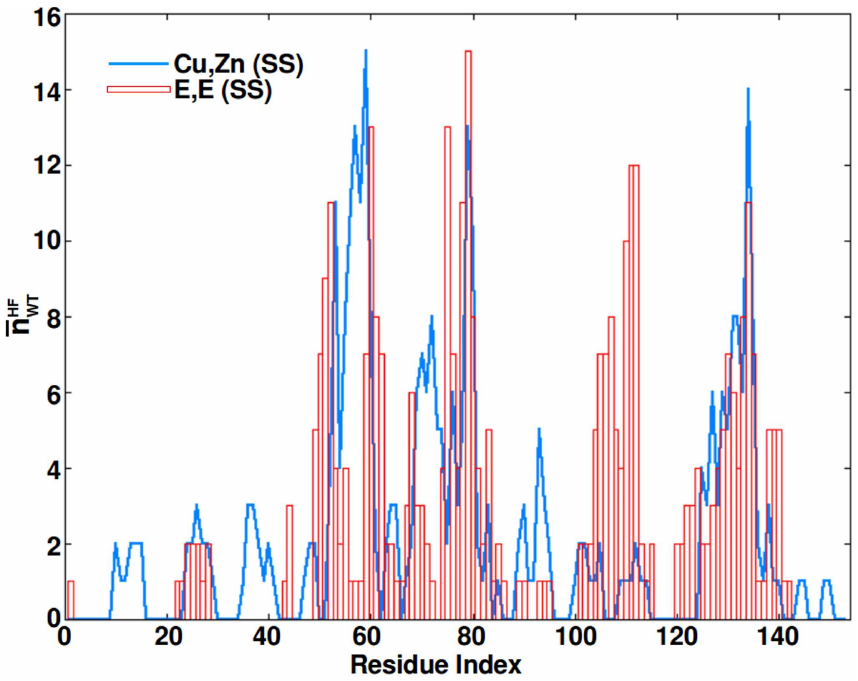}}
\caption{Mean number of highly frustrated contacts in the Cu,Zn(SS) and E,E(SS)
  states of WT SOD1, as a function of
  residue index.}
\label{figapohf}
\end{figure}

The 30 residues mutated to alanine are: 5,10,15,20,25,30,
35,40,45,50,56,59,65,70,75,80,85,91,96,100,105,110,115, 
120,125,
130,135,139,144 and 150.  
We again took 50 snapshots from the equilibrium simulations of each
protein to calculate the ensemble average 
of highly frustrated contacts $\overline{n}^{\tiny HF}$.
We performed frustratometer analysis for each
Cu,Zn(SS) or E,E(SS) alanine mutant, and calculated the difference in
number of highly frustrated residues between WT SOD1 and the alanine mutant
as a function of sequence index $i$: $\overline{n}^{\tiny HF}_{\tiny
  X\rightarrow A} (i) - \overline{n}^{\tiny HF}_{\tiny WT} (i)$.
We then averaged this quantity over all alanine mutants, 
$\left<\overline{n}^{\tiny HF}_{\tiny
  X\rightarrow A} (i)\right>_X - \overline{n}^{\tiny HF}_{\tiny WT}
(i) \equiv
\left<\overline{n}^{\tiny HF} \right>_{\tiny MUT} - \overline{n}^{\tiny HF}_{\tiny WT}$,
and plotted the result in Figure~\ref{fignonals}, panel A as a
function of sequence index. 

Upon mutation to alanine at the sequence locations above, the number of highly frustrated
contacts increased on average by 9 for Cu,Zn(SS) SOD1 and decreased on
average by 29 for E,E(SS) SOD1. This indicates that mutations
which normally increase frustration in the Cu,Zn(SS) state of WT SOD1
generically decrease frustration in the WT E,E(SS) state.
This recapitulates the findings in the main text for ALS-associated
mutations, and supports the conclusion that the E,E(SS) state is frustrated. 

\renewcommand{\thefigure}{S\arabic{figure}}
\renewcommand{\arraystretch}{1}
\begin{figure}[h]
\centerline{\includegraphics[width=0.5\textwidth]{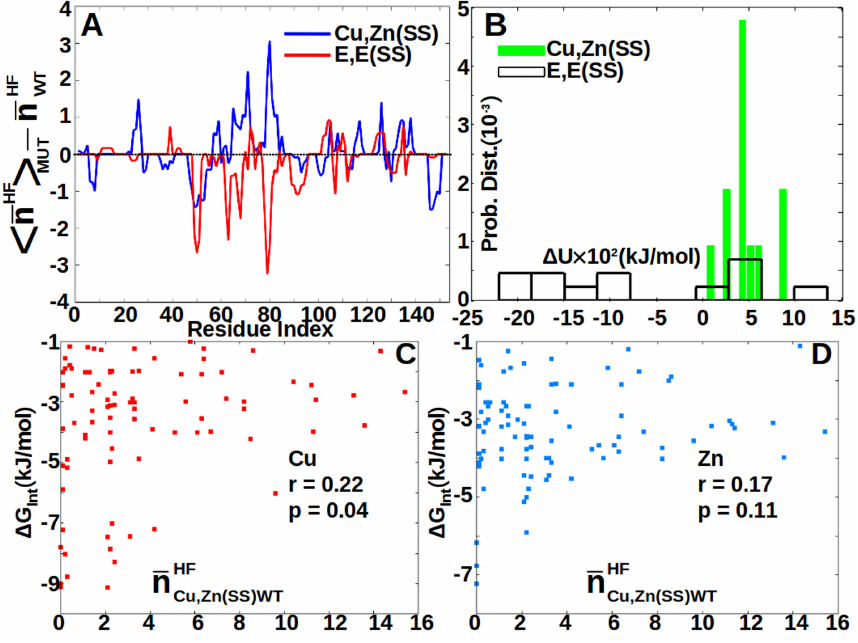}}
\caption{
(Panel A) The number of frustrated contacts within a sphere of radius
5\AA~centered on each C${}_\alpha$ atom, is found as a function of residue
index, $n^{\mbox{{\tiny HF}}}(i)$. Ensemble averages are taken from 50 snapshots in an equilibrium
simulation to obtain $\overline{n}^{\mbox{{\tiny HF}}}(i)$. This is done for both the Cu,Zn(SS) state and the E,E(SS)
state, for both the  WT sequence, and for 30 alanine
mutants (Table~S3). The 30 alanine mutant sequences are averaged to
obtain the ensemble and mutant averaged number of contacts as a
function of residue index $i$, $\left< \overline{n}^{\mbox{{\tiny
        HF}}}(i)\right>_{\mbox{{\tiny MUT}}}$. Plotted is the
difference between $\left< \overline{n}^{\mbox{{\tiny
        HF}}}(i)\right>_{\mbox{{\tiny MUT}}}$ and the corresponding
numbers for the WT sequence $\overline{n}_{\mbox{{\tiny
      WT}}}^{\mbox{{\tiny HF}}}(i)$ .
A positive
number would indicate an increase in frustration upon mutation. Cu,Zn(SS)
state is shown in blue and has an average of +9 contacts. E,E(SS) state is
shown in red and has an average of -29 contacts.
(Panel B) Distribution of the thermal average potential energy
change upon mutation, $\Delta \overline{U}_{{\mbox{\tiny X}}
  \rightarrow {\mbox{\tiny A}}}$, for the 30 alanine mutants given in
the text. Distributions for both the
Cu,Zn(SS) state and the E,E(SS) state of SOD1 are shown.
(Panel C) Scatter plot of the interaction free energy of a given residue's side
chain with Cu, {\it vs.} the mean equilibrium number of highly frustrated contacts
that residue has in the Cu,Zn(SS) state of WT SOD1. The scatter plot shows
essentially no correlation.
(Panel D) Same as Panel C, but for the interaction free energy with
Zn. The scatter plot again shows no correlation.
}
\label{fignonals}
\end{figure}

\subsection{Potential energy analysis of a non-ALS related alanine scan of SOD1}
\label{nonalsU}

We have also calculated the difference in
potential energy upon mutation, for ALS-associated mutants in the main
text, and the 30 alanine mutants described above (Table~S3). 
The procedure for calculating the potential energy change upon
mutation is described in the Methods. 
The potential energies, averaged over 50 equilibrium conformations,
are obtained both before and after mutation, and the difference
obtained as $\Delta \overline{U}_{{\mbox{\tiny X}}
  \rightarrow {\mbox{\tiny A}}}$. This difference is obtained in both
the Cu,Zn(SS) and E,E(SS) forms of SOD1. 
Plotted  in Figure~\ref{fignonals}, panel B are the distributions of 
$\Delta \overline{U}_{{\mbox{\tiny X}}
  \rightarrow {\mbox{\tiny A}}}$ over the above 30 non-ALS associated
alanine mutations, for the both the Cu,Zn(SS) and E,E(SS) forms of
SOD1. The same quantities are plotted in Figure~4B of the
main text for 22 ALS-associated mutations. 
In Figure~\ref{fignonals}, panel B, we see that all 30 mutants raise (penalize)
the potential energy of Cu,Zn(SS) SOD1, while most of the 30 mutants
lower the potential energy of E,E(SS) SOD1.
These findings are consistent with those in the main text, and
indicate that the results that obtained for ALS-associated mutants
are also common to non-ALS SOD1 mutants, and so are a generic feature
of SOD1.

\subsection{Correlation between metal interactions and frustration,
  and metal interactions and distance}
\label{sectintfrus}

Frustratometer analysis of the equilibrium E,E(SS) state of WT SOD1 shows
that 90 residues have at least 1 highly-frustrated contact
(Fig.~\ref{figapohf}). The mean number of highly-frustrated contacts of
each of these residues,
$\overline{n}_{\mbox{{\tiny E,E(SS)}}}^{\mbox{{\tiny HF}}}$, may be
compared with the interaction free energy of that residue with the
metal $\Delta G_{int}$. The interaction free energy is obtained from
metal extraction as described in the main text and in the Methods
below. This analysis directly
tests the correlation 
between the degree of frustration of a residue, and
that residue's involvement in binding metals. 
Figure~3, panels B and C of the main text show a strong correlation
between metal interaction free energy and the frustration of a
particular residue in the E,E(SS) state: the more frustration a
residue has, the stronger its interaction with the metals. 
As a control test, the metal interaction free energy of a given residue may be
compared with its number of highly-frustrated contacts in the
Cu,Zn(SS) state. Here we see no correlation for either Cu or Zn
(Fig.~\ref{fignonals}, panels C,D). Hence, on a residue by residue
basis, frustration in the apo
state of SOD1 facilitates metal affinity for both Cu and Zn. 

\renewcommand{\thefigure}{S\arabic{figure}}
\renewcommand{\arraystretch}{1}
\begin{figure}[h]
\centerline{\includegraphics[width=0.5\textwidth]{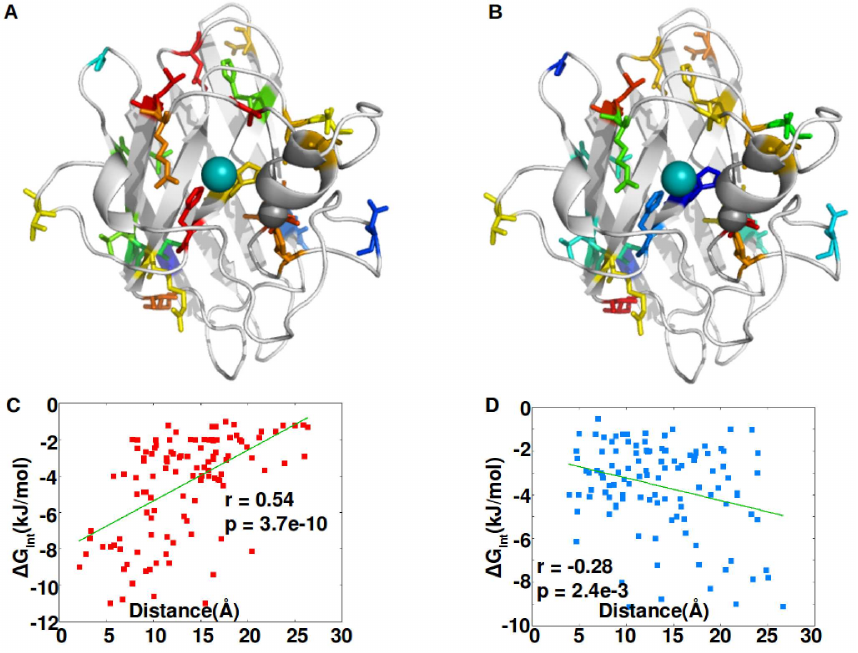}}
\caption{
Panel (A): Residues color-coded by interaction energy with the Cu ion
(depicted as a cyan sphere). The extent
of interaction is strongest in magnitude for red colored residues and
decreases to blue.
Panel (B): same as (A) for the Zn ion (depicted as a grey sphere).
Panel (C): Interaction energy with Cu correlates with the distance of the residue from the Cu
ion;  residues in close proximity more strongly interact with the metal.
Panel (D): Interaction energy with Zn does not correlate with distance
of the residue to the Zn ion, indicating non-local allosteric
effects. }
\label{figCuZncorr}
\end{figure}

The distance-dependence of the interaction free energy between a
residue and either Cu or Zn was investigated in Figure~5 of the main
text. Results were shown for the 24 mutants listed in
Table~\ref{tabmutants}. 
When the dataset corresponding to the 90 residues with at least one
highly frustrated contact is used (Table~\ref{tabmutants}), 
the general result remains true that allosteric regulation
for metal affinity is significantly correlated with the proximity of a
residue to the Cu binding site, but not with the proximity of a
residue to the Zn binding site (Fig.~\ref{figCuZncorr}). 
There is more scatter with this larger dataset (but also selective to
only include frustrated residues) in the plot measuring correlation
between interaction free energy and distance to the Cu, and the
correlation decreases somewhat from that in the main text. 
The conclusion remains true that the affinity
for Zn is imposed collectively across the whole protein.

\subsection{Relationship between mechanical work, fluctuations, potential
energy, and frustration}
\label{sectcompare}

In Figure~\ref{figcompare}, we plot the cumulative work distribution,
RMSF, time-resolved change in potential energy after mutation, and
frustratometer results for the ALS-associated mutants studied here:
G127X and A4V. 

\renewcommand{\thefigure}{S\arabic{figure}}
\renewcommand{\arraystretch}{1}
\begin{figure}[h]
\centerline{\includegraphics[width=0.5\textwidth]{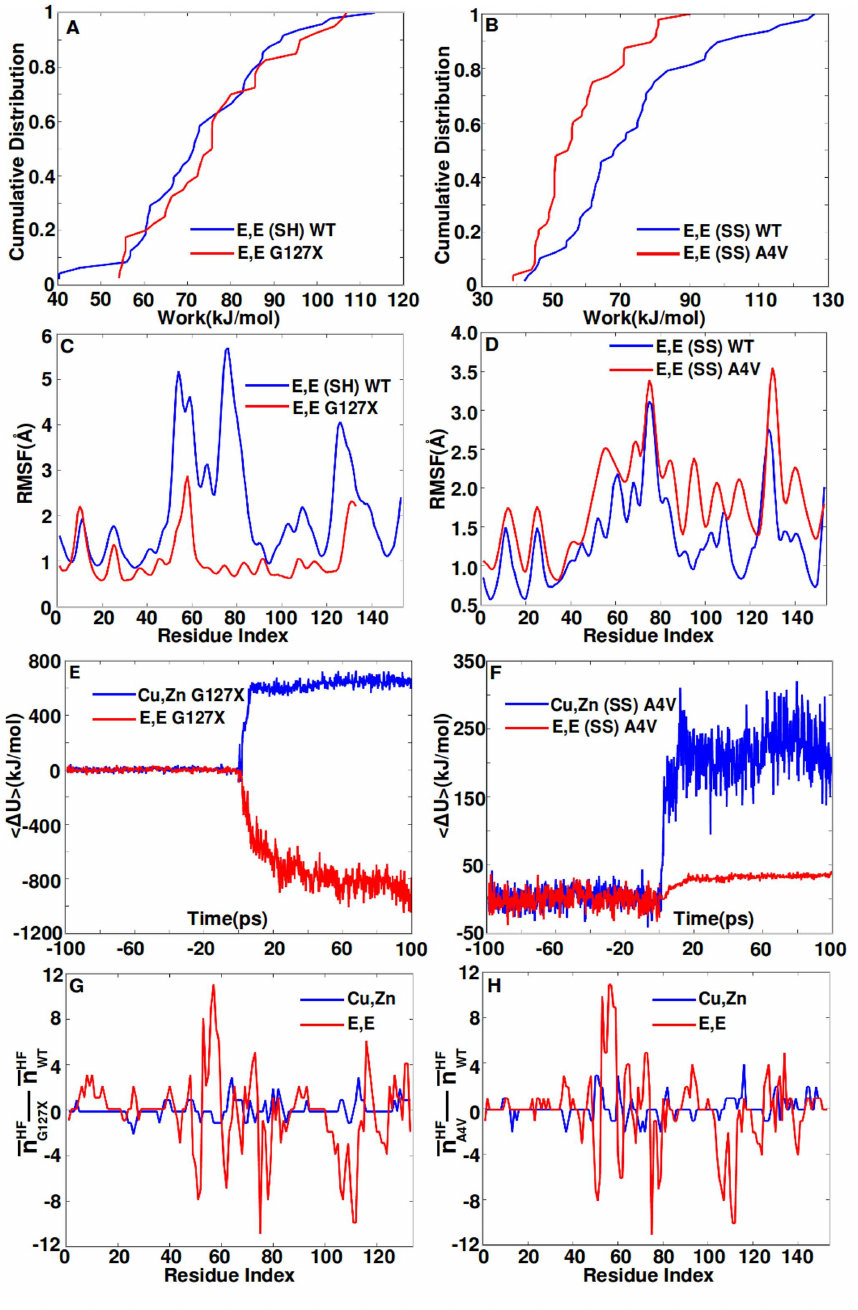}}
\caption{Panel (A): Cumulative distribution of work values to pull
  residues to 5\AA, for E,E G127X and E,E(SH) WT. Panel (B): same as
  (A), for E,E(SS) A4V and E,E(SS) WT. Panels (C,D): RMSF values as
  function of residue index, for the same proteins in Panels
  (A,B). Panels (E,F): time-resolved change in total potential energy
  after mutation, from E,E(SH) WT $\rightarrow$ E,E G127X and Cu,Zn(SH) WT
  $\rightarrow$ Cu,Zn G127X (panel E) and
  from E,E(SS) WT $\rightarrow$ E,E(SS) A4V and Cu,Zn(SS) WT
  $\rightarrow$ Cu,Zn(SS) A4V (panel F).
Panel (G): Difference in the number of highly-frustrated contacts for
each residue, G127X - WT, for Cu,Zn(SH) and E,E(SH) variants. Panel
(H): Same difference as in (G), for A4V - WT, for Cu,Zn(SS) and
E,E(SS) variants.
}
\label{figcompare}
\end{figure}

These two mutants show different behavior: The cumulative distribution
of work values shows that E,E G127X is stabilized with respect to
E,E(SH) WT (Fig.~1B in the main text and Fig.~\ref{figcompare}A). The
mechanical stability of A4V shows the opposite effect 
however (Fig.s~\ref{figmetalss} and~\ref{figcompare}B): the
mutation softens the protein. 

The dynamic fluctuations are consistent with the work-value analysis:
the fluctuations are reduced upon mutation for G127X
(Fig.2C in main text and Fig.~ref{figcompare}C), but enhanced for A4V
(Fig.~ref{figcompare}D).  Likewise, the total potential energy in the
apo protein 
decreases upon mutation for G127X (Fig.~\ref{figcompare}E) but
increases upon mutation for A4V (Fig.~\ref{figcompare}F).
The potential energy increases in the holo protein for both mutants. 

The frustratometer analysis indicates that the frustration increases
in the holo protein for both mutants, 5 contacts for G127X and 4 contacts for
A4V. However for the apo proteins, the number of highly frustrated contacts in E,E G127X
decreases from E,E(SH) WT by 16 contacts on average, and the
number of highly frustrated contacts in E,E G127X decreases from that
in E,E(SS) WT by 35 contacts. For 
E,E(SS) A4V, frustration is decreased from that in E,E(SS) WT by 31
contacts. 
Thus frustration decreases from the E,E(SS) state slightly more in
G127X than in E,E(SS) A4V. 
A general analysis of the decrease in potential energy and frustration
for apo mutants, along with the results from their individual
mechanical scans, is an interesting topic for future work.

\section{Methods}
\label{methods}

\subsection{Steered molecular dynamics simulations}
\label{sectsteeredmd}
Steered molecular dynamics
(SMD) using constant-velocity moving restraints was used to simulate
the action of moving AFM cantilever on a protein.
The general procedure used two tethering points for the pulling simulations. 
In our study of SOD1 monomer, one tether was placed at the position
of the alpha carbon closest to the center of mass of the protein
(generally $C_\alpha (46)$), and another tether was placed on the
alpha carbon of a particular amino acid to be pulled on.
To implement a mechanical stability scan across the surface of the
protein, every 5th amino acid along with the first and last residues was
selected for a pulling simulation. This coarse-grains the mechanical
profile, so that the work values obtained are a sampling of the
work values for all residues.

SMD
simulations~\cite{SotomayorM2007,Carrion-VazquezM2003,IrbackA2005,ImparatoA2008}
were preformed in GROMACS to monitor the forced
unfolding of SOD1 variants.
An all-atom representation of the protein was used, with
OPLS-AA/L parameters~\cite{JorgensenWL96,KaminskiG2001}, and
a generalized Born surface area (GBSA) implicit solvent model was
used, with dielectric constants of the protein and solvent taken
to be 4 and 80 respectively. The Onufriev-Bashford-Case (OBC)
algorithm was used to calculate Born radii~\cite{OnufrievA2004}.
The phenomenological surface tension coefficient used in calculating
solvation free energies was
0.005 kcal/mol/(\AA$^2$)~\cite{BjelkmarP2010}.
The LINCS algorithm was applied to constrain all bond lengths that contained
a hydrogen atom~\cite{HessB1997}.

Prior to simulation, energy was minimized
using a steepest descent algorithm to remove any
potential steric clashes.
Simulations were carried out with an integration time-step of 2 fs and coordinates were saved after every 100 ps.
The long-distance cut-off used for non-bonded interactions was 14 \AA~
for both Electrostatic and van der Waals (VDW) interactions.
A cubic simulation box with periodic boundary conditions was used,
with size such that all protein atoms were initially at least 20 \AA~ from any cubic face.
For every SOD1 variant protein considered, a simulation was run for 20 ns to equilibrate the
protein before
any pulling simulations were performed and data was collected.

Because only moderate deformations were required to construct
mechanical force-extension curves, a very slow pulling speed (for
simulations) of
$2.5\times 10^{-3}$m/s was admissible. This speed is comparable to those
used in AFM experiments~\cite{LvS2010}, and is  $400-8000\times$
slower than the speeds generally used in simulations~\cite{LemkulJ2010,KimT2006,GeriniM2003}.
The spring constant of the simulated AFM cantilever was taken to be
$5$ kJ/mol/\AA$^2$.
The average $F(x)/x$ values from the epitope pulling
simulations e.g. for residues 10 and 17, in the extension range of 1-5
\AA, are 5.1 and 7.3 kJ/mol/\AA$^2$ respectively,
indicating that the protein effective spring constant is comparable to
the cantilever spring constant.
The average modulus as determined by $2 W(x)/x^2$ between 1 and 5 \AA~
is $7.72$ and $10.35$ kJ/mol/\AA$^2$ for residues 10 and 17
respectively.
The average modulus between 0 and 1 \AA~ is much higher: $67.08$ for
residue 10 and $91.46$ for residue 17 in the same units.
Each pulling simulation ran until the change in
distance between the two tethering points was 5 \AA.

\subsection{Explicit solvent simulations}

Pulling simulations of WT SOD1 were also performed in explicit water. Equilibrated
starting structures were generated by placing the protein in a cubic
box of simple point charge (SPC) water,
such that all protein atoms were initially at least 20 \AA~from any
cubic face.
This required  $10866$ water molecules to solvate the protein. Prior
to simulation, the energy of the system was minimized using a steepest descent algorithm.
The system was then equilibrated for $20$ ns
under a constant volume (NVT) ensemble, with temperature maintained
at 300 K using the Berendsen weak coupling method.
Pulling simulations were then carried out using an integration
time-step of 2 fs,  and coordinates were saved every 100 ps.
The long-distance cut-off used for non-bonded electrostatic and VDW interactions was 14 \AA.

We have also minimized and equilibrated E,E (SS) WT, E,E (SH) WT and E,E G127X
variants of SOD1 in explicit solvent
following the same procedure as described above for Cu,Zn (SS) WT SOD1.  Each
of the variants was equilibrated for 35 ns, and the last 20 ns of that
trajectory was used for the calculation of RMSF and backbone amide
SASA.

\subsection{Tethering residues}
\label{secttethering}

The  $C_\alpha$ atom of the residues listed in Table~\ref{tabtether}
were closest to the center of mass of the corresponding protein
variant, and so were used as the central tether.
Whenever the pulling residue appeared to be very close
to the center of mass tethering residue, the tethering residue was
shifted to the midpoint of the protein sequence to avoid artificially
large forces.
Specifically, for several variants listed in Table~\ref{tabtether}, residues 40, 45, 46, 50, 54 and 55 were
either too close to the central $\Ca$, or were along the same beta
strand as the central $\Ca$, and thus gave anomalous force-extension profiles probing covalent
bonding topology moreso than non-covalent stabilizing
interactions. Thus
when residues 40, 45, 46, 50, 54 and 55 were pulled, the tethering residue
was moved to corresponding residue indicated in
Table~\ref{tabtether}.
This always resolved the problem of large forces, except for 1-110
SOD1. For this variant, we have taken residue 86, which is also close
to the center of mass, because the midpoint residue (56) did not
resolve the problem of anomalously large forces.



\subsection{Residues used for mechanical scans of protein stability}
\label{secttethers}

We have previously developed a model calculation of the free energy of
unfolding specific regions of a protein using the so-called
single-sequence approximation~\cite{GCPalgorithm09}, in analogy to
similar models used in protein
folding~\cite{PlotkinSS96,HilserVJ96,Munoz99:PNAS,AlmE99,FinkelsteinAV99:pnas}. In
this formalism, the resulting free energy landscape is constructed as
a function of the center residue of the contiguous strand taken to be
disordered, and the sequence length of the disordered strand.  In our
unfolding model, we have treated electrostatic and van der Waals
energies in the system at the level of the
CHARMM22 energy function~\cite{GuestW10bcb,GuestW11pccp}. Protein entropy was calculated from
simulations of the unfolded ensemble in explicit solvent.  Regions of
low thermodynamic stability may be projected onto the native
structure; these regions constitute candidate epitopes for
unfolding-specific immunologic therapy,
providing that the criteria of residue-specific immunogenicity and
uniqueness of the epitope sequence in the proteome are established.
Antibodies that bind to disease-specific protein epitopes
corresponding to non-native conformations displayed by disordered
protein sequences have been developed for superoxide dismutate 1
(SOD1)~\cite{RakhitR07} and Prion protein~\cite{ParamithiotisE03}.
The candidate epitopes as determined from free energy landscape
analysis for WT SOD1 are given in the table in Figure~\ref{figtabepitopes}.

Every 5th residue along the protein sequence, including the first and
last residues,  was taken as a tethering residue. As well,
the central
positions of the predicted epitopes and anti-epitopes were added to
the data set.

The first anti-epitope contains 5 residues with residue $18$ at its
center, however the free energy landscape prediction gave a larger
stability to candidate epitope 1 than epitope 2~\cite{GCPalgorithm09},
so we chose residue $17$ instead as potentially more representative
for the stiff sequence of residues. For all other epitopes or
anti-epitopes, either the center residue was chosen as the tethering
point if the stand contains an odd number of residues, or if the
strand contained an even number of residues, the pulling residue was
chosen randomly between the two center residues.

If the center of the epitope happened to coincide with
a multiple of 5 already included in the scan (e.g. residue 10), then
the next residue (residue 11 in this case) was also taken. Thus
a set of $48$ residues was used in the mechanical
scans: (1, 5, 10, 11,
15, 17, 20, 24, 25, 30, 31, 35, 38, 40, 45, 46, 50, 54, 55, 60,
65, 70, 73, 75, 80, 85, 90, 91, 95, 100, 101, 105, 108, 110, 115, 120,
121, 125, 130, 135, 136, 140, 141, 145, 146, 150, 151, 153).

\subsection{Proteins considered}
Proteins whose mechanical profiles were calculated are given in
  Table~\ref{tabtether} below, along with
the corresponding PDB entries used to generate equilibrated structures for
simulations. 


\subsection{Modeling proteins with no PDB structure}
\label{modelingnopdb}
Some proteins studied here have no PDB structural coordinates
available; see Table~\ref{tabtether}. For these proteins, a structure
was built by modifying known PDB structures of similar proteins.  
For example, metal-depleted (E,E (SS)) WT SOD1 was created from E,E (SS) SOD1
structure(1RK7) by back-mutating Q133E, E50F and E51G using the PyMOL
software package~\cite{DeLanoW2002}.  Metal-depleted,
disulfide-reduced (E,E (SH) WT) SOD1 was created from both E,E (SS) SOD1 structure
(1RK7) by reducing the disulphide linkage and mutating E50F, E51G, and
Q133E, and also from Cu,Zn (SS) WT SOD1 (1HL5) by reducing the disulphide
linkage and removing the metals.
The E,E G127X mutant was created from Cu,Zn (SS) WT SOD1 by removing the metals and the last 20
residues, and then mutating the last 6 residues of the remaining
sequence (KGGNEE) to correspond to the frame-shifted non-native
C-terminal peptide sequence (GGQRWK) prior to the termination
sequence. The Cu,Zn G127X mutant was prepared following the same way as done for
E,E G127X - only the metals were kept intact.
The full-length C57S/C146S mutants were prepared from Cu,Zn (SS) WT SOD1 by mutating
both Cys-57 and Cys-146 to Serine (Cu,Zn C57S/C146S 1-153) and simultaneously removing
the metals (E,E C57S/C146S 1-153). The truncated versions of these variants were made
by following the same technique as described for full-length proteins - only residues
128-153 were deleted for the truncated ones. 
E,E (SH) SOD1 variants 1-140 and 1-110 were
created from the full length E,E (SH) WT SOD1 by deleting the
last 13 and 43 residues respectively.  As described in the
Methods subsection on steered MD simulations, all modified structures were energy
minimized and equilibrated for 20 ns by running an equilibrium
simulation, before performing any pulling assays.


\subsection{G\={o} model energetics}
\label{goenergetics}
The SMOG G\={o} model recipe used here~\cite{WhitfordP09} takes heavy atoms
within $2.5$ \AA, and applies native contacts to them with a Lennard-Jones 6-12
potential. The SMOG G\={o} model recipe also attributes energies to
dihedral terms in the BB and SCs. The energy scale determining
well-depths for these interactions is given through the relation: $N_c
\epsilon_c + N_{BB} \epsilon_{BB} + N_{SC} \epsilon_{SC} = N_A \cdot (1
\mbox{kJ/mol})$, where $N_c$ is the number of non-local contacts and
$\epsilon_c$ is their corresponding well-depth, and the other numbers
correspond to BB and SC dihedral numbers and energy scales
respectively. $N_A$ on the right hand side is the number of atoms. The
overall energy scale of all interactions is thus given by $1$kJ/mol
times the number of atoms in the system.

\subsection{Umbrella Sampling and Weighted Histogram Analysis Method (WHAM)}
\label{whammethodmech}

Initial configurations were obtained from pulling
simulations as described in the Methods subsection on steered MD, to obtain $25$ initial
conditions between $0$ and $5$\AA. These initial configurations are then
simulated for $10$ns each, in an umbrella potential with
stiffness $500$kJ/mol/nm$^2$ to constrain the simulations
near their corresponding
separation distances. The $25$ simulations are then used to
reconstruct the free energy profile along the distance coordinate
using the weighted histogram analysis method (WHAM)~\cite{KumarS1992},
and the free energy difference between $0$ and $5$\AA~is then
obtained.

\subsection{Metal insertion or extraction in calculating interaction energies}
\label{whammetal}

To obtain free energies of metal binding,  metals were
inserted into or extracted from the putative binding locations for WT SOD1, and all
glycine mutants of SOD1. Metals were always found to be at least
metastable in their putative binding positions.
To obtain the metal binding free energy, the
most direct straight-line path was first determined for the pulling
direction, by finding the approximate direction of highest solvent
exposure of the metal. Tethers were placed on the metal, and on the
closest C$_\alpha$ atom opposite to the direction of highest solvent
exposure. 
For Cu extraction this corresponded to tethering the C$_\alpha$ atom of
residue Phe45,  and for Zn
removal the corresponding C$_\alpha$ atom tether was in Asp83.

The metal was pulled away from the protein using a spring constant of $500$
kJ/mol/nm$^2$ and a pull rate of $0.01$ nm/ps.
For purposes of the free energy calculation, the final extension from
the equilibrium distance between metal and tethering residue
was taken to be approximately $30$\AA.
From these trajectories, snapshots were taken to generate the starting
configurations for the umbrella
sampling windows~\cite{PateyG1973,TorrieG1974,TorrieG1977}.
An asymmetric distribution of
sampling windows was used, such that the window spacing was
1\AA~between 0 and 20 \AA~separation, and 2\AA~beyond 20\AA~of
separation. Such
spacing allowed for increasing detail at smaller separation distances, and
resulted in a total of 25 windows. In each window, 10 ns of MD was performed for
a total simulation time of 250 ns utilized for umbrella
sampling. Analysis of the results was performed using the weighted
histogram analysis method (WHAM)~\cite{KumarS1992}.

To remove conformational distortion effects on the free energy,
position restraints were applied to the protein in the following way. For a given
$C_\alpha$ atom in the protein, all other $C_\alpha$ atoms within
$5$\AA~were constrained to have a roughly constant $C_\alpha$-$C_\alpha$
distance, the same as in an equilibrated structure, using spring
constants of $392\times 10^3$ kJ/mol/nm${}^2$.
Using $C_\alpha$ constraints allows the protein
to retain its structure under force,
while still allowing the side chains and partially the backbone to
fluctuate in response to the external perturbation. After metal
extraction, $C_\alpha$ constraints were relaxed and
the relaxation free energy was calculated as described below.

The free energy for metal extraction was corrected
to account for protein relaxation in the final metal depleted
state: $\Delta F_{tot} = \Delta F_{constrained}^{extraction} + \Delta
F_{relax}$. That is, after metal extraction,
$C_\alpha$ constraints are gradually reduced from $392\times 10^3$
kJ/mol/nm${}^2$ to zero using 30 windows and 10ns of relaxation time
in each window, and the free energy change for this process is again
obtained using WHAM.
Convergence of
the free energy values was tested by both varying the number of windows
(to 30, 35 and 45) and varying the length of the equilibrium
simulation in each window (to 15, 20, 25 and 30 ns). In all cases the
free energies were seen to have converged using the original
protocol.

To validate that equilibrium free energies have been obtained by this procedure, thermodynamic
cycles were constructed by re-inserting the metal and equilibrating,
for WT SOD1, and ALS-associated mutants A4V and G93A, as well as PTM variants
Cu,E(SS) and E,Zn(SS)~\cite{DasA12jmb}. The metal binding free
energies subject to $C_\alpha$ constraints were calculated, as well as
the subsequent relaxation free energies. The total free energy change
over such cycles was typically less than $0.04$ kJ/mol, confirming
that the cycles are thermodynamic to within the error of the simulation.
A more detailed description of the procedure is given in
reference~\cite{DasA12jmb}. 

\subsection{Potential energy calculations}
\label{secpot}

Cu,Zn(SS) and E,E(SS) WT SOD1 protein structures were simulated in a cubic
box with periodic boundary conditions and size such that all protein atoms were
initially at least 20 \AA~ from any cubic face. An all-atom representation of the
protein was used, with OPLS-AA/L parameters~\cite{JorgensenWL96,KaminskiG2001}, and
a generalized Born surface area (GBSA) implicit solvent model~\cite{BjelkmarP2010} was used, with
dielectric constants of the protein and solvent taken to be 4 and 80 respectively.
The Onufriev-Bashford-Case (OBC) algorithm was used to calculate Born radii~\cite{OnufrievA2004}.
The phenomenological surface tension coefficient used in calculating solvation free
energies was 0.005 kcal/mol/(\AA$^2$)~\cite{BjelkmarP2010}. The LINCS algorithm was
applied to constrain all bond lengths that contained a hydrogen atom~\cite{HessB1997}.
Prior to simulation, energy was minimized using a steepest descent algorithm to remove
any potential steric clashes. Simulations were carried out with an integration time-step
of 2 fs and coordinates were saved after every 100 ps. The long-distance cut-off used
for non-bonded interactions was 14 \AA~ for both Electrostatic and van der Waals (VDW)
interactions. 

For both the Cu,Zn(SS) and E,E(SS) WT SOD1 variants of SOD1 protein, a simulation was run for
20 ns to equilibrate the protein. Once equilibrium is reached, the
potential energy was recorded and averaged over 50 snapshots, where
each snapshot is taken every $0.5$ns. 
The final equilibrated conformation
was then mutated in PyMol to either one of the ALS-associated mutants
described in the main text or alanine mutants described above. 
The mutated structures were then again equilibrated until convergence
of the potential energy (typically $\approx 35$ns).
Then 50 snapshots were taken, once every $0.5$ns, and these snapshots
were used to obtain the ensemble-averaged potential energy of the
mutant. In this way, $\Delta \overline{U}_{\mbox{\tiny WT} \rightarrow
  \mbox{\tiny MUT}}$ is obtained for each mutant. A histogram of this
quantity is plotted for the 30 non-ALS alanine mutants described in
the above text, in Figure~\ref{fignonals}B.

\subsection{Frustratometer analysis}
\label{sectfrus}

Highly frustrated contacts were found using the ``frustratometer'' method of Ferreiro
{\it et. al.}, given in reference~\cite{FerreiroDU11}. This method
finds the particular contacts that are highly frustrated in a one protein
conformation. Metal-protein interactions are not included in the
frustratometer analysis; only protein-protein interactions are
considered. We have used equilibrium protein conformations of the Cu,Zn(SS) and E,E(SS)
states of SOD1 to distinguish the effect of metals on SOD1 protein structure. 
An equilibrium ensemble of 50 conformations was obtained for both WT SOD1
and mutants of SOD1, as described in the previous section above.  
The number of highly frustrated contacts within a sphere of
5\AA~around each C$_\alpha$ atom was
found for every snapshot, and averaged over snapshots. This number was
again averaged over mutants to obtain the mutant and equilibrium
ensemble-averaged number of highly frustrated contacts around a given
C$_\alpha$ atom.

\subsection{Statistical Analysis}
\label{statanal}

\renewcommand{\thefigure}{S\arabic{figure}}
\renewcommand{\arraystretch}{1}
\begin{figure}[h]
\centerline{\includegraphics[width=0.5\textwidth]{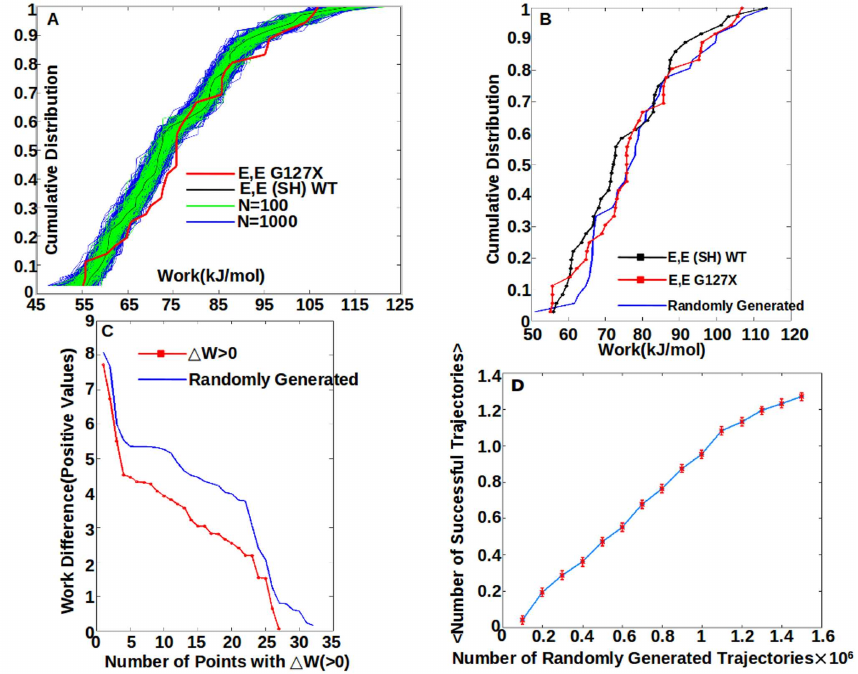}}
\caption{
A Statistical significance test between E,E (SH) WT SOD1 and E,E G127X
cumulative distributions. The black curve denotes the E,E (SH) WT
profile, and the red curve denotes the E,E G127X
profile. The collection of green lines corresponds to 100
randomly-generated cumulative distributions in the following way:
starting from the E,E (SH) WT work values, gaussian noise with zero
mean and standard deviation $\sigma = 2.7$~kJ/mol  is added to
each work value. The resulting collection of work values are then
rank-ordered to generate a cumulative distribution, and this process
is then repeated $N$ times. The collection of blue lines
corresponds to $N=1000$. For large enough $N$, one will begin to find
outlying cumulative distributions which deviate as much or more than a
given trial distribution, here E,E G127X. An example of
such an outlier is shown in panel B.
In this case, the number of positive deviations (differences in work
value) between the red and
black curves in panel A are recorded (27), along with their work
values, which are themselves rank ordered largest to smallest and
plotted in panel C. An outlier at least as extreme as
E,E G127X must have a set of positive work deviations
such that the corresponding  curve lies above the red curve in panel
C. In this case the randomly generated outlier has 32 positive
deviations in work, 27 of which are larger than those between
E,E G127X and E,E (SH) WT variants.
The number of randomly generated trajectories is increased, until the
mean number of successful trajectories, corresponding to the criterion
in panel C, exceeds unity. A plot of the mean number of successful
trajectories as a function of $N$ is given in panel D. The
statistical significance $p$ of a given trajectory is then given as
$p=1/N_1$, where $N_1$ is the number of randomly generated
trajectories that give an expectation value of unity for the number of
successfully-generated outliers.
In this case, the statistical significance of E,E G127X
is about $1/(1.1\times 10^6)$, or about $9\times 10^{-7}$.
}
\label{figsstest}
\end{figure}

The mechanical profile is a collection of $48$
work values for a given SOD1 variant.
Each work value measured from our simulations is accurate to some
$\sigma$ kJ/mol which we determine as follows.
The difference in the mechanical scans of two initial
conditions, for example as derived from different crystal structures,
gives a measure of the error in the mechanical scan. For
E,E (SH) WT SOD1 shown in the main panel of Figure~\ref{figStructinde}, the work
difference $\Delta W$ has mean $0.03$kJ/mol and standard deviation
$\sigma_o = 2.1$kJ/mol. A z-test indicates that with
$93\%$ certainty the values of $\Delta W$ arise from a gaussian
distribution of mean zero and variance $\sigma_o$. On the other hand,
the values of $\Delta W$ in the inset A of Figure~\ref{figStructinde} for Cu,Zn (SS) WT SOD1 have
mean $0.75$kJ/mol and standard deviation $1.5$kJ/mol. This
distribution fails a z-test for a gaussian with mean zero and
$\sigma_o=1.5$kJ/mol, indicating that the error in measurement is
larger than $\sigma_o$. We increase the standard deviation until the
z-test indicates the data arises from a gaussian of mean zero and
standard deviation $\sigma$. This gives $\sigma=2.7$kJ/mol, which we
use as a measure of the error bars in the work values of the
mechanical profile.

In comparing two mechanical
profiles residue by residue, the probability of finding the second
profile by chance, given the measurements $W_i$ of the first profile,
and the error bars $\sigma$, is:
\begin{equation}
P\left ( \left\{ \Delta W_i\right\}\right )= \frac{1}{2} \prod_{i=1}^N
\mbox{erfc}\left(\frac{ \left| \Delta W_i \right|}{\sqrt{2} \sigma}
\right)  \: .
\label{eqstatsig1}
\end{equation}
where $\left| \Delta W_i \right|$ is the modulus of the difference
between the two work values for residue $i$, and $N=48$ here.

Comparing two cumulative distributions is more forgiving, in that there is a
greater likelihood two different mechanical profiles may still give
the same overall work distribution. These two protein variants might
then be said to have the same overall mechanical stability, although
what parts of the structure were the most or least stable would differ
for each protein.

Because the overall variance of a given work profile is so large however
(typically $\approx 15$~kJ/mol), it is very likely to obtain the
cumulative distribution of any one variant from any other variant by
a direct Kolmogorov-Smirnov test. We are interested in a different
question however: given one work profile, its cumulative distribution, and the error bars associated with
the work values, what is the probability of obtaining a
cumulative distributions at least as extreme as another given one by chance,
i.e. as arising from the original work profile?
This is a complicated question because it is non-trivial to account
for the combinatorial number of ways of obtaining a given cumulative
distribution.

To find the above probability, and thus determine the
statistical significance of a given cumulative distribution, we employ
a ``Monte Carlo'' procedure of generating cumulative distributions from
a given ``baseline'' work profile.
For example, to test the effect of truncation at the C-terminus
which mechanically stabilizes the protein, we wish to find the probability
of obtaining the E,E G127X cumulative distribution (or a
distribution even more stabilized) from the E,E (SH) WT SOD1 cumulative
distribution. The E,E G127X cumulative distribution has
$30$ work values that are more stable than those in the E,E (SH) WT SOD1
cumulative distribution; the difference in work values are plotted in
Figure~\ref{figsstest} panel C,
largest to smallest. We seek cumulative distributions that have rank
ordered deviations at least as large as those of the
E,E G127X cumulative distribution.

Work profiles are constructed by
adding gaussian noise to a the baseline work profile (here that of E,E (SH) WT
SOD1), where the
gaussian noise has mean zero and standard deviation
$\sigma=2.7$~kJ/mol. A cumulative distribution is then constructed for
each generated profile by sorting the values lowest to highest. After
constructing $N$ of such cumulative distributions, we ask whether one
has found a cumulative distribution with, e.g. rank-ordered positive
deviations at least as large as those in the E,E G127X
cumulative distribution. The value of N where the expected number of
trajectories is unity determines the statistical significance:
$p=1/N$.
Plots of the corresponding distributions for E,E (SH) WT SOD1 and
E,E G127X SOD1 are given in Figure~\ref{figsstest}.

If 2 cumulative distributions are different, their mechanical 
fingerprints are necessarily different. If they are the same, their 
fingerprints can always be compared to discern the effects of mutation. 
In practice however we never encountered two different fingerprints that 
gave rise to indistinguishable cumulative distributions.


\end{article}

\section{Supporting Information Tables}
\label{tables}

\renewcommand{\thetable}{S\arabic{table}}
\renewcommand{\arraystretch}{1}
\begin{table}[H]
\centering
\caption{\bf{Correlation table of Work(upper diagonal) and RMSF(lower diagonal) values for different variants
of SOD1}}
\begin{tabular}{c|c|c|c|c|c|c}
\hline
{RMSF}/{Work} & Cu,Zn (SS) WT & E,E (SS) WT & E,E (SH) WT & E,E G127X & E,E 1-140 & E,E 1-110 \\
\hline
Cu,Zn (SS) WT & {1}/{1} & -0.37 & -0.16 & 0.20 & 0.06 & 0.14 \\
\hline
E,E (SS) WT & 0.16 & {1}/{1} & 0.51 & -0.17 & -0.03 & 0.22 \\
\hline
E,E (SH) WT & 0.10 & 0.51 & {1}/{1} & -0.09 & -0.13 & 0.38 \\
\hline
E,E G127X & 0.06 & -0.14 & 0.13 & {1}/{1} & 0.32 & 0.62 \\
\hline
E,E 1-140 & 0.02 & 0.09 & 0.11 & 0.21 & {1}/{1} & 0.56 \\
\hline
E,E 1-110 & 0.03 & 0.13 & 0.11 & 0.21 & 0.22 & {1}/{1} \\
\hline
\end{tabular}
\label{tabcorr}
\begin{flushleft}
{\small Correlations are generally insignificant. The two significant
  correlations are between E,E G127X and E,E 1-110 (p=1.0e-4), and E,E 1-140 and E,E
  1-110 (p=6.2e-4).}
\end{flushleft}
\end{table}

\renewcommand{\thetable}{S\arabic{table}}
\renewcommand{\arraystretch}{1}
\begin{table}[H]
\centering
\caption{
\bf{Central tethering residues for mechanical profiles}}
\begin{tabular}{ccc}
\hline
SOD1 variant & Tethering residue & PDB ID used for \\
 & & structure generation \\
\hline
Cu,Zn (SS) WT & 46 (76${}^\ddag$) & 1HL5, 2C9V \\
E,E (SS) WT & 46 (76${}^\ddag$) & 1RK7${}^\dag$ \\
E,E (SH) WT & 46 (76${}^\ddag$) & 1HL5, 1RK7${}^\dag$ \\
Cu,Zn G127X & 45 (67${}^\ddag$) & 1HL5${}^\ast$ \\
E,E G127X & 45 (67${}^\ddag$) & 1HL5${}^\ast$ \\
E,E 1-140 & 46 (71${}^\ddag$) & 1HL5 \\
E,E 1-110 & 46 (86${}^\ddag$) & 1HL5 \\
Cu,Zn C57S/C146S 1-153 & 46 (76${}^\ddag$) & 1HL5${}^\oplus$ \\
E,E C57S/C146S 1-153 & 46 (76${}^\ddag$) & 1HL5${}^\oplus$ \\
Cu,Zn C57S 1-127 & 45 (64${}^\ddag$) & 1HL5${}^\ominus$ \\
E,E C57S 1-127 & 45 (64${}^\ddag$) & 1HL5${}^\ominus$ \\
\hline
\end{tabular}
\begin{flushleft}
${}^\ddag$for residue 40, 45, 46, 50, 54 and 55. \\
${}^\dag$Structures reported have mutations from the WT sequence; these were ``back-mutated''
to construct structures for the WT sequence, as described in the
Methods subsection on modeling proteins with no PDB structure. \\
${}^\ast$residues 134-153 deleted and 128-133 mutated. \\
${}^\oplus$Cys-57 and Cys-146 mutated to Serine. \\
${}^\ominus$Cys-57 mutated to Serine and residues 128-153 deleted.
\end{flushleft}
\label{tabtether}
\end{table}

\renewcommand{\thetable}{S\arabic{table}}
\renewcommand{\arraystretch}{1}
\begin{table}[H]
\centering
\caption{
\bf{Proteins considered for Frustratometer results and Interaction Energy calculations}}
\begin{tabular}{ccc}
\hline
Frustratometer & Interaction Energy \\
\hline
A4V,G37R,L38V,G41D,G41S,H43R, & A4G,L8G,I17G,L38G,H43G,H46G, \\
H46R,H46R/H48Q,T54R,D76Y,H80R, & H46G/H48G,T54G,D76G,H80G,D83G,L84G, \\
L84F,G85R,D90A,G93A,G93C,E100G, & D90G,E100G,I113G,R115G,D124G,D125G, \\
I113T,D124V,D125H,S134N,L144F${}^\ddag$ & S134G,A140G,R143G,L144G,V148G,I149G${}^\dag$ \\
\hline
V5A,G10A,Q15A,F20A,S25A, & A1G,P13G,E21G,Q22G,K23G,E24G,S25G,N26G,P28G,V31G,S34G, \\
K30A,I35A,E40A,F45A,F50A, & E40G,E49G,F50G,D52G,N53G,A55G,C57G,T58G,S59G,A60G,P62G, \\
G56A,S59A,N65A,K70A,K75A, & H63G,F64G,N65G,P66G,L67G,S68G,R69G,K70G,H71G,P74G,K75G, \\
H80A,G85A,K91A,D96A,E100A, & E77G,E78G,R79G,V81G,N86G,A89G,V94G,A95G,D101G,S102G,V103G, \\
A105A,H110A,R115A,H120A,D125A, & I104G,S105G,L106G,S107G,D109G,H110G,C111G,I112G,H120G,E121G,K122G, \\
G130A,T135A,N139A,L144A,G150A${}^\ast$ & A123G,L126G,K128G,N131G,E132G,E133G,T135G,K136G,T137G,N139G,S142G${}^\oplus$ \\
\hline
\end{tabular}
\begin{flushleft}
${}^\ddag$Proteins used for Figure 3. \\
${}^\dag$Proteins used for Figure 5, Figure S18 and Figure S19. \\
${}^\ast$Proteins used for Figure S18. \\
${}^\oplus$Proteins used for Figure S18 and Figure S19.
\end{flushleft}
\label{tabmutants}
\end{table}

\renewcommand{\thetable}{S\arabic{table}}
\renewcommand{\arraystretch}{1}
\begin{table}[H]
\centering
\tiny
\caption{{\small \bf{Mechanical work values of WT SOD1 variants - the values reported in the table
depict the work in kJ/mol needed to pull a particular residue up to 5 \AA. ${}^{\dag}$}}}
\begin{tabular}{cccccccccccc}
\hline
Residue & Cu,Zn & E,E & E,E & Cu,Zn & E,E & Cu,Zn 1-153 & E,E 1-153 & Cu,Zn 1-127 & E,E 1-127 & E,E & E,E \\
Index & (SS) WT & (SS) WT & (SH) WT & G127X & G127X & C57S/C146S & C57S/C146S & C57S & C57S &1-140 & 1-110 \\
\hline
1   &    66.32     &    63.90     &    60.32     &    62.47     &    60.37     &    67.12     &    61.25     &    64.62     &    66.37     &    61.09     &    54.08     \\
5   &    68.85     &    74.92     &    88.95     &    71.83     &    75.80     &    76.02     &    81.27     &    70.66     &    72.22     &    66.40     &    68.90     \\
10  &    66.18     &    94.68     &    81.29     &    82.38     &    86.41     &    92.94     &    72.22     &    75.83     &    61.27     &    70.19     &    72.00     \\
11  &    108.41     &    62.53     &    59.51     &    68.39     &    65.69     &    70.20     &    66.80     &    64.41     &    80.27     &    64.09     &    58.71     \\
15  &    89.97     &    71.56     &    68.81     &    99.48     &    104.00     &    77.44     &    91.12     &    75.83     &    69.37     &    88.09     &    76.67     \\
17  &    91.53     &    61.60     &    45.15     &    99.48     &    95.69     &    73.91     &    66.87     &    72.83     &    72.27     &    94.09     &    67.84     \\
20  &    78.24     &    67.77     &    60.67     &    67.49     &    64.74     &    87.41     &    82.12     &    76.87     &    90.29     &    92.40     &    65.87     \\
24  &    58.82     &    54.46     &    56.77     &    57.99     &    55.69     &    67.01     &    72.22     &    69.32     &    81.29     &    64.09     &    52.77     \\
25  &    62.31     &    126.04     &    87.35     &    70.39     &    72.65     &    73.22     &    78.18     &    65.93     &    77.22     &    74.69     &    69.98     \\
30  &    42.90     &    124.00     &    113.40     &    51.00     &    54.39     &    76.48     &    59.16     &    84.37     &    68.22     &    73.50     &    72.93     \\
31  &    66.38     &    76.52     &    55.92     &    81.20     &    85.69     &    67.20     &    71.29     &    65.02     &    69.22     &    74.09     &    65.59     \\
35  &    77.33     &    104.09     &    84.16     &    79.39     &    76.61     &    66.55     &    69.27     &    76.93     &    72.12     &    62.59     &    67.33     \\
38  &    67.11     &    63.59     &    64.74     &    79.43     &    75.69     &    76.01     &    80.18     &    70.77     &    78.20     &    54.09     &    57.77     \\
40  &    94.41     &    98.00     &    91.57     &    75.48     &    73.67     &    68.33     &    90.19     &    87.83     &    71.22     &    59.20     &    64.95     \\
45  &    83.49     &    64.31     &    68.18     &    67.88     &    65.04     &    76.04     &    81.20     &    64.83     &    61.22     &    76.90     &    61.17     \\
46  &    63.32     &    56.23     &    66.83     &    79.97     &    75.69     &    66.12     &    71.27     &    74.44     &    60.27     &    84.09     &    67.54     \\
50  &    57.08     &    112.86     &    78.24     &    71.98     &    73.14     &    73.51     &    69.37     &    70.73     &    59.38     &    67.59     &    63.63     \\
54  &    85.03     &    51.12     &    71.63     &    59.99     &    55.69     &    60.20     &    61.11     &    60.84     &    67.37     &    64.09     &    59.27     \\
55  &    88.66     &    78.95     &    72.65     &    94.99     &    95.16     &    69.05     &    58.27     &    59.86     &    78.22     &    80.09     &    76.12     \\
60  &    79.50     &    79.77     &    58.41     &    65.00     &    69.98     &    75.93     &    66.27     &    65.37     &    81.29     &    91.19     &    66.11     \\
65  &    92.99     &    76.29     &    83.28     &    88.99     &    87.99     &    86.94     &    71.27     &    66.83     &    61.29     &    66.19     &    72.21     \\
70  &    80.56     &    64.42     &    95.86     &    68.34     &    69.06     &    66.07     &    58.33     &    77.79     &    58.48     &    94.30     &    73.95     \\
73  &    125.06     &    43.42     &    60.82     &    107.09     &    105.69     &    70.20     &    71.22     &    94.93     &    69.32     &    104.09     &    78.50     \\
75  &    72.82     &    94.37     &    71.42     &    59.49     &    55.12     &    78.13     &    65.37     &    75.94     &    66.33     &    69.90     &    58.25     \\
80  &    73.18     &    42.31     &    72.12     &    98.40     &    99.67     &    68.05     &    80.33     &    73.90     &    78.22     &    72.09     &    72.58     \\
85  &    72.10     &    77.37     &    70.86     &    56.49     &    54.14     &    66.76     &    66.76     &    72.08     &    61.22     &    85.40     &    64.01     \\
90  &    87.98     &    62.90     &    74.42     &    78.49     &    80.03     &    83.32     &    71.22     &    63.88     &    69.44     &    59.40     &    62.38     \\
91  &    98.44     &    46.27     &    61.36     &    58.40     &    55.69     &    74.91     &    90.44     &    76.04     &    82.22     &    54.09     &    53.91     \\
95  &    61.74     &    90.15     &    82.99     &    97.88     &    96.10     &    84.16     &    62.22     &    85.33     &    90.19     &    101.20     &    81.48     \\
100 &    97.82     &    81.67     &    87.44     &    57.39     &    54.90     &    73.11     &    71.27     &    70.93     &    81.22     &    78.19     &    71.92     \\
101 &    65.32     &    75.58     &    40.28     &    82.39     &    85.69     &    74.20     &    66.66     &    75.93     &    78.79     &    84.09     &    66.99     \\
105 &    73.50     &    71.67     &    63.50     &    65.30     &    62.21     &    80.68     &    77.72     &    82.02     &    71.22     &    71.90     &    61.38     \\
108 &    69.29     &    54.35     &    40.29     &    77.00     &    75.69     &    66.12     &    81.18     &    70.04     &    80.12     &    64.09     &    56.23     \\
110 &    66.24     &    96.08     &    86.64     &    82.47     &    79.07     &    73.55     &    60.38     &    65.88     &    71.29     &    96.50     &    55.02  \\
115 &    81.13     &    57.64     &    72.76     &    101.09     &    106.83     &    76.08     &    71.29     &    83.90     &    66.65     &    61.59     &    *    \\
120 &    81.67     &    58.15     &    101.18     &    87.30     &    85.65     &    81.57     &    59.33     &    66.02     &    61.20     &    81.19     &    *    \\
121 &    113.67     &    59.55     &    66.77     &    78.02     &    75.69     &    70.44     &    56.48     &    85.00     &    59.43     &    84.09     &    *    \\
125 &    64.12     &    83.58     &    82.85     &    76.00     &    77.76     &    70.37     &    61.22     &    65.32     &    66.22     &    71.30     &    *    \\
127 &    -             &    -             &    -             &    -             &    -             &    -             &    -             &    64.02    &    70.69   &    -             &    *    \\
130 &    87.92     &    68.13     &    103.14     &    71.32     &    72.34     &    86.75     &    71.27     &    *            &    *           &    55.59     &    *    \\
133 &    -             &    -             &    -             &    68.03    &    66.35 &    -             &    -             &    *  &    *   &    -             &    *    \\
135 &    79.41     &    116.09     &    85.13     &    *            &    *         &    96.52     &    80.69     &    *  &    *   &    62.29     &    *    \\
136 &    62.52     &    46.75     &    70.28     &    *  &    *      &    67.12     &    71.22     &    *  &    *   &    74.09     &    *    \\
140 &    99.75     &    77.40     &    92.17     &    *  &    *      &    74.94     &    66.66     &    *  &    *   &    57.70    &    *    \\
141 &    147.35     &    58.33     &    56.82     &    *  &    *      &    64.76     &    71.22     &    *  &    *   &    *  &    *    \\
145 &    72.82     &    62.23     &    76.21     &    *  &    *      &    66.20     &    60.16     &    *  &    *   &    *  &    *    \\
146 &    80.53     &    45.31     &    61.20     &    *  &    *      &    67.12     &    61.29     &    *  &    *   &    *  &    *    \\
150 &    68.12     &    74.84     &    80.22     &    *  &    *      &    74.76     &    66.44     &    *  &    *   &    *  &    *    \\
151 &    76.67     &    61.72     &    66.17     &    *  &    *      &    73.83     &    61.27     &    *  &    *   &    *  &    *    \\
153 &    69.05     &    69.50     &    60.34     &    *  &    *      &    66.12     &    60.33     &    *  &    *   &    *  &    *    \\
\hline
\label{tabworkvals}
\end{tabular}
\begin{flushleft}
{\small ${}^\dag$  ``$\ast$'' indicates residues that are not present in the given
variant, while ``$-$'' indicates work values that were not measured for
a given variant.}
\end{flushleft}
\end{table}